%% file: Zero_Coupon_Rate.tex
\documentclass[12pt]{article}

\usepackage{amsfonts}
\usepackage{graphicx}

\newcommand\fiverm{\tiny\rm}
\input  ./prepictex
\input  ./pictex
\input  ./postpictex

\begin{document}
\baselineskip1.8em

\setcounter{page}{1}

\begin{center}
\textbf{\large\bf The Zero-Coupon Rate Model for Derivatives Pricing }
        \\[1cm] Xiao Lin\footnote{E-mail: xiao\_lin\_99@yahoo.com; Finished: 2016-05-23; Last Revised: 2022-02-23.}  
        \\[5mm]
       {\small Global Market Department \\[2mm] \small Industrial and Commercial Bank of China \\ \small Beijing 100140, China }

\vspace{5mm}
\end{center}

\hspace{8mm} {\begin{minipage}{13cm}
\baselineskip1.7em
{\sc Abstract}:
	The aim of this paper is to present a dual-term structure model of interest rate derivatives in order to solve the two hardest problems in financial modeling: the exact volatility calibration of the entire swaption matrix, and the calculation of bucket vegas for structured products. The model takes a series of long-term zero-coupon rates as basic state variables that are driven directly by one or more Brownian motion. The model volatility is assigned in a matrix form with two terms. A numerical scheme for implementing the model has been developed in the paper. At the end, several examples have been given for the model calibration, the structured products pricing and the calculation of bucket vegas.
\vspace{5mm}


\end{minipage}}

\vspace{5mm}

\section{\large Introduction}
\setcounter{equation}{0}

The pricing of interest rate options and structured products depends on a mathematical model, which formulates the dynamics of the term-structure of interest rates. Many sophisticated models of term-structure interest rates have been developed and used for the derivatives trading business, which we can divide into three major groups.  The first group is the short rate model, which assumes that the instantaneous spot rate follows a stochastic process. There are many one-factor short rate models available, such as the Vasicek model [1], CIR model [2], Ho-Lee model [3], Hull-White model [4], Black-Derman-Toy (BDT) model [5], and the Black-Karasinski (BK) model [6]. There are also some multi-factor short-rate models such as the two-factor Longstaff-Schwartz model [7]. The second group of interest rate models is the forward rate model, such as the Heath-Jarrow-Morton (HJM) model [9] and the Brace-Gatarek-Musiela (BGM) model [10], which assume that a series of forward rates (either instantaneous forward rates or 3-month forward rates) follows a stochastic process. The third group of rate models is the long-term spot rate model, like the Markov Functional model proposed by Hunt, Kennedy and Pelsser [12], in which, through a pricing formula of digital swaption, the swap rate is the state variable driven by a stochastic process.
	In the past decade, a large amount of structured products (e.g., callable CMS spread swap) appeared on the market. The fact that different models would give products very different prices has been a big puzzle to market participants. Usually, when a model is proposed for pricing structured products in a bank, the model validation department in the bank would ask if the model:
\begin{enumerate} \baselineskip1.2em
\item[(A)] is arbitrage-free;
\item[(B)] can be calibrated to the basic swaption or cap/floor instruments exactly;
\item[(C)] can be used to calculate the risk factors (include bucket vegas) accurately.
\end{enumerate}
Unfortunately, the existing models listed above have not been able to give a satisfactory answer to all three requirements. Specifically, one of the problems is the global calibration of each model to the entire at-the-money swaption volatility matrix. Since a structured product depends on a set of uncertain market indexes, using different swaption instruments to calibrate a model will give different prices. Certainly, the model validation department would not believe there are more than two risk-free prices for one deal. So far, the calibrations by any models (including a study by Johnson [11] using the MF model) are still approximate. The later developed BGM model had more modeling factors, which brought more flexibility to the calibration. However, there was always a great deal of debate on the best global calibration of the entire at-the-money swaption matrix, but we still cannot find a satisfactory report on the task. The second problem is the calculation of bucket vegas to each volatility component. Bucket vegas are very important to banks since they are the basic data for hedging products. For a simple structure like the basic Bermudan swaption, the short rate model is used to calculate bucket vegas by specifically picking up a series of volatility instruments along an anti-diagonal line. However, this method failed in pricing generic structured products linked to indexes of constant-maturity swap (CMS).

In this paper, we present a new model for interest rate dynamics to approach all three model validation requirements. The main advantages of the model are: 1) The model takes a series of long-term zero-coupon rates as basic state variables. The zero-coupon rates directly represent the market index of CMSs, which are most frequently used in the structured products. Thus, this feature gives us a best view in pricing the CMS structured products. 2) It is a dual-term structure model of interest rates, one term being the modeling time at which the zero rate is observed, the other being the life length of the zero rate. These two terms in the model volatility space correspond with the two terms of the swaption volatility matrix. This feature enables us to exactly calibrate the model to the entire swaption matrix, and calculate bucket vegas for structured products.

\section{\large The model}
\setcounter{equation}{0}

The fundamental assumptions for the model are similar to those of most models, i.e, (i) interest rate products like zero-coupon bonds, swaps, swaptions, and structured products are liquidly traded in the  market; (ii) all transaction costs, e.g. bid/ask spreads and tax, etc., can be ignored; (iii) the market information is transparent, such that there exist no arbitrage opportunities. Like other models, we assume (iv) all market prices of interest rate products are driven by one or more financial quantities. However, the specific driving factors in our model are the zero rates of constant-maturity bonds, which will be described shortly below.

In this section, we will develop an interest rate model in a continuous market, in which a stochastic interest rate process, say $r = r(t)$,  are assumed as continuous functions of time $t$. In the Appendix we will extended the model to a simple market where interest rate process becomes right continuous.

Let $t$ and $T$ be two time steps with $0\leq t<T$, and $P(t, T)$ be the time $t$ price of a zero-coupon bond that matures at time $T$. We know the bond has a life length of $T-t$  at time $t$. In our work, we will model the zero coupon rate (in this paper, we call it zero rate in short) of the bond, denoted by $y=y(t, T-t)$, in which we specify the bond life term $T-t$ as an argument, emphasizing it as the rate of a constant-maturity bond. One kind of instruments corresponding to the zero rate is the CMSs which are widely traded in the market. Thus, the bond price can be expressed by
\begin{eqnarray} \label{2.1}
		P(t,T)=\exp\Big\{-y(t, T-t)\cdot(T-t)\Big\}.
\end{eqnarray}

We denote the bank account value at time $t$ by
\begin{eqnarray} \label{2.2}
		B(t)=\exp\Big\{\int_0^tr(\tau)d\tau\Big\},
\end{eqnarray}
where $r(t)$ is a stochastic short rate process. Let us consider a bond with very short term $T=t+h$. If we invest $\$1$ to the bond at time $t$, during the period  $h$, we get interest  $(1-P(t, t+h))/P(t,t+h)\approx y(t, h)\cdot h$. If we do the same in the bank account we get interest approximated to $r(t)\cdot h$. In the non-arbitrage market, these two interests should be same. Thus, we assume that $\lim_{T\to t} y(t, T-t) = r(t)$, and need not to study the stochastic property for $r(t)$ separately. At following we will show that the short rate is a bridge liked to two zero rates of the same life lengths at different times.

Further we define a forward rate. Let $y(t, T-t)$  and $y(t, T-t+h)$ be two zero rates at time $t$, for two bonds matured on  $T$ and $T+h$, respectively. The time-$t$ forward rate for the interval $T\leq t<T+h$ is defined by
\begin{eqnarray}	\label{2.3}
		f(t, T-t, h) & = & \frac{1}{h}\log\frac{P(t, T)}{P(t, T+h)}\cr \nonumber \\
					 & = & \frac{y(t, T-t+h)\cdot (T-t+h)-y(t, T-t)\cdot (T-t)}{h}.
\end{eqnarray}
Using the forward rate, the bond price $P(t, T+h)$  can be expressed as
\begin{eqnarray} \label{2.4}
		 P(t, T+h)=\exp\Big\{-y(t, T-t)\cdot(T-t)-f(t, T-t, h)\cdot h\Big\},
\end{eqnarray}
which implies the interest earned by $P(t, T+h)$  can be separated as two parts, one is from the zero rate of an early matured bond  $P(t, T)$, and the other is from forward rate. The forward rates is another bridge between two bonds with different maturities.

Now we come to the fundamental pricing theory. It states that in a non-arbitrage market, there exists a risk-neutral measure, with the bank account as numeraire, so that the ratio
\begin{eqnarray} \label{a2.5}
		\frac{P(t, T)}{B(t)}=\exp\Big\{-y(t, T-t)\cdot (T-t)-\int_0^tr(\tau)d\tau\Big\}
\end{eqnarray}
is a martingale for time period $0\leq t\leq T$.  Thus, applying the Ito-Doeblin formula to (\ref{a2.5}),
\begin{equation} \label{a2.6}
		{\rm d} \bigg[ \frac{P(t,T)}{B(t)} \bigg] = e^{(\cdots)} \bigg[ -(T-t){\rm d}y + y{\rm d} t - r{\rm d} t
				+\frac{1}{2} (T-t)^2 {\rm d}y^2 \bigg],
\end{equation}
the martingale condition means the right-hand side of the above equation is proportional to a Brownian motion  d$w_t$. We assume
\begin{equation} \label{a2.7}
		-(T-t){\rm d}y + y{\rm d} t - r{\rm d} t +\frac{1}{2} (T-t)^2 {\rm d}y^2 = -(T-t)\sigma \, {\rm d} w_t,
\end{equation}
where $\sigma = \sigma(t,T-t)$  is the normal volatility associated to  $y=y(t,T-t)$. Take a variance to the above equation and ignore the high-order term of d$t$  we have  ${\rm d}y^2 = \sigma^2{\rm d}t$. Thus, (\ref{a2.7}) becomes
\begin{equation} \label{a2.8}
		(T-t){\rm d}y - y{\rm d}t + r{\rm d}t -\frac{1}{2} (T-t)^2 \sigma^2 {\rm d}t = (T-t)\sigma \, {\rm d} w_t.
\end{equation}
Let $s=T-t$, ${\rm d}s = -{\rm d}t$, since
\begin{eqnarray}
		& &(T-t){\rm d}y -y{\rm d}t \ = \ s \bigg[ y(t+{\rm d}t,s + {\rm d}s) - y(t,s) \bigg] + y(t,s)\, {\rm d}s
			 \nonumber\\[2mm]
			&=& s \bigg[ y(t+{\rm d}t,s + {\rm d}s) -y(t+{\rm d}t,s) + y(t+{\rm d}t,s) - y(t,s) \bigg] + y(t,s)\, {\rm d}s
			 \nonumber\\[2mm]
			&=& s\bigg[ y(t+{\rm d}t,s) - y(t,s) \bigg] + s \bigg[ y(t+{\rm d}t,\xi + {\rm d}s) -y(t+{\rm d}t,s) \bigg] + y(t,s)\, {\rm d}s.\nonumber\\
\end{eqnarray}
We use $\partial y = y(t+{\rm d}t,s) - y(t,s)$ to denote the partial derivative of  $y=y(t,s)$ with respect to the first argument $t$, in which the second argument $s = T-t$  is kept unchanged. Therefore,
\begin{equation}
			s\bigg[ y(t+{\rm d}t,s) - y(t,s) \bigg] =  s \, \partial y = (T-t)\, \partial y(t,T-t).
\end{equation}
Also, we denote the instantaneous forward rate of the bond $P(t,T-t)$ by  $f(t,T-t,0)$.  Then,
\begin{eqnarray}
			& & s \Big[ y(t+{\rm d}t ,s + {\rm d}s) - y(t+{\rm d}t,s) \Big] + y(t,s)\, {\rm d}s
				 \nonumber\\[2mm]
			&=& s \Big[ y(t,s + {\rm d}s) -y(t,s) +{\mathcal O}({\rm d}t^2) \Big] + y(t,s +{\rm d}s)\, {\rm d}s + {\mathcal O}({\rm d}t^2)
				 \nonumber\\[2mm]
			&=& y(t,s+{\rm d}s) \cdot(s+{\rm d}s) - y(t,s)\cdot s +  {\mathcal O}({\rm d}t^2)
				  \nonumber \\[2mm]
				&=& -f(t,T-t,0) {\rm d}t +  {\mathcal O} ({\rm d}t^2).
\end{eqnarray}
After dropping the term of ${\mathcal O}({\rm d}t^2)$, (\ref{a2.8}) becomes
\begin{equation}
		(T-t){\partial}y - (f-r){\rm d}t -\frac{1}{2} (T-t)^2 \sigma^2 {\rm d}t = (T-t) \sigma \, {\rm d} w_t.
\end{equation}
Thus, devided by $(T-t)$ in the formula, we obtain the dynamic equation of the normal process for the zero rate $y(t,T-t)$,
\begin{equation} \label{SDE}
		{\partial}y \, = \, \Bigg[ \frac{f-r}{T-t} +\frac{1}{2} (T-t) \sigma^2 \Bigg] {\rm d}t + \sigma \, {\rm d} w_t.
\end{equation}

Remark: The model has formulated a normal process for the zero rate dynamics.  A lognormal model process can be obtained when the normal volatility $\sigma$ in (\ref{SDE}) is replaced by lognormal volatility $y\sigma$.  However, we should keep in mind a problem of negative interest rates in the recent EUR, JPY and CHF markets. Some models failed in the pricing due to their fundamental assumption of the lognormal distribution on the financial variables.

In order to model the dynamics of a term-structure interest rate, we consider a set of zero-coupon bonds
$P(t,T_1), P(t,T_2), \cdots, P(t,T_K)$ that have life terms $s_1 = T_1 - t$, $s_2 = T_2-t$, $\cdots$, $s_K = T_K-t$.
The $k$-th zero rate  is represented by $y_k(t) = y(t,s_k)$, and the same is for the forward rate  $f_k(t) = f(t,s_k,0)$.   In the framework of a one-factor model, we assume all zero rates are driven by one Brownian motion d$w_t$ with different volatility $\sigma_k(t) = \sigma(t,s_k)$. When we apply (\ref{SDE}) to these $K$ zero rates, we have a set of $K$ dynamic equations:
\begin{equation} \label{2.24}
			\partial y_k(t) \,= \, \Bigg\{   \frac{ f_k(t) - r(t) } {s_k} + \frac{1}{2} \sigma_k^2(t)\, s_k   \Bigg\}
			{\rm d}t+\sigma_k(t)\, {\rm d} w_t, \hspace{5mm} k = 1,2,\cdots,K.
\end{equation}
When $K$ zero rates $y_k(t) = y(t,s_k)$, $k = 1,2, \cdots, K$  are found, we will be able to build a discount factor curve at time $t$ for pricing derivatives.

The dynamic equations can be extended to a multi-factor model in the following way. Let d$\bold{w} = ({\rm d}w_a, {\rm d}w_b)$   be a two-dimensional independent Brownian motion. The diffusion term in (\ref{2.24}) can be written as
\begin{equation}   \label{2.25}
		\langle \vec{\bf \sigma}_k \cdot {\rm d}{\bf w} \rangle =
		\sigma_k \, (\cos\theta \cdot {\rm d}w_a + \sin\theta \cdot {\rm d} w_b).
\end{equation}			 					
We can take $\theta$  as the function of the rate life $s_k$:
\begin{equation}   \label{2.26}
		\theta = \frac{\pi}{2}\frac{s_k}{L_{\max}},		
\end{equation}			 					
where  $L_{\max}$ is the maximum life length of the indexes, e.g., the maximum years of swap terms in the swaption matrix. Thus, the short-term rate $y_1$  mainly depends on d$w_a$, while the long-term rate $y_K$   will follow  d$w_b$. This should be a good setup for modeling CMS spread options (e.g. CMS30Y$-$CMS2Y), if the long-term and short-term rates are independent of each other. Similarly, we can make a three-dimensional Brownian motion as
\begin{equation}   \label{2.27}
		\langle \vec{\bf \sigma}_k \cdot {\rm d}{\bf w} \rangle =
		\sigma_k \, (\cos\theta \cos\phi \cdot {\rm d}w_a + \sin\theta \cos\phi \cdot {\rm d} w_b + \sin\phi \cdot {\rm d}w_c).
\end{equation}			 					
where $\phi$ can be a function of time $t$. Thus the general form of term-structure dynamics becomes
\begin{equation} \label{2.28}
			\partial y_k(t) \,= \, \Bigg\{   \frac{ f_k(t) - r(t) } {s_k} + \frac{1}{2} \sigma_k^2(t)\, s_k   \Bigg\} {\rm d}t
			+  \langle \vec{\bf \sigma}_k \cdot {\rm d}{\bf w} \rangle, \hspace{5mm} k = 1,2,\cdots,K.
\end{equation}

At the end of this section, we have a remark on the  model volatility $\sigma = \sigma(t,s)$.
It also has two arguments which are corresponding to those of zero rate $y(t,s)$.
Thus, this model is a true dual-term structure model.
In the numerical computation, the model volatility surface $\sigma (t,s)$  can be  represented
by finite discrete parameters $\sigma_{ik} = \sigma (t_i,s_k)$  at some term points.
Values at non-term points will be obtained by a bi-linear interpolation.

\section{\large Building stochastic process}
\setcounter{equation}{0}

Beginning from this section, we present a numerical scheme for implementing the dual-term zero rate model. We will focus on those topics specifically with the zero rate model. Other common pricing techniques will not be discussed.

Let $t=t^0, t^1, t^2, \cdots$  be a partition of time with finite small steps $h=t^{n+1}-t^n$. First, we define a zero-coupon discount factor for modeling. Let the probability distribution of a zero rate $y(t^n,s)$ at time $t^n$ with life term $s$ is represented by a set of value $y(t^n,s,q)$, $q = 1,2,\cdots, Q$. The number $q$ in the expression can also be referred to the index of a Monte Carlothe path. The zero-coupon discount factor of $y(t^n,s,q)$ is defined by
\begin{equation} \label{3.1}
	{\rm df}^n(s,q) = \exp\Big\{-y(t^n,s,q))\cdot s \Big\}.
\end{equation}
We call it the discount factor for short in this paper.  A state at $t^n$ of a path is a yield curve that is represented by a series of discrete discount factors ${\rm df}^n(s_k,q)$, $k = 1,2,\cdots,K$, together with an interpolation method. We use the discount factor curve as the state variable because it is the most basic element for calculating many other financial quantities. However, since the stochastic factor in dynamic equation is the zero rate of the bond, we need to change the zero rate into the discount factor back and forth constantly in building the stochastic probability field.

\input {./fig_301.tex}

Figure~1 is a sketch of one modeling path. The modeling starts at valuation date ($n=0$), where the discount factor curve is known. Without loss of generality, let us assume that we have already built up the curves at step $n$, denoted by  ${\rm df}^n(s,q)$, where $q$ stands for the path,  and $s\ge 0$  is the curve term. For simplicity, we will temporarily drop the path index $q$ for a while, and denote the discount factor curve as  ${\rm df}^n(s)$.  Now we want to build a curve on step  $n+1$. First, from the time step length  $h=t^{n+1}-t^n$, we can get a discount factor ${\rm df}^n(h)$  at $s=h$  from this known curve by an interpolation, which corresponds to a deterministic short rate at $t^n$  by
\begin{equation} \label{3.2}
		r^n = \frac{1}{h}\, \ln \frac{{\rm df}^n(0)}{{\rm df}^n(h)}.
\end{equation}
Thus, we obtain the first discount factor at the bottom point of step $t^{n+1}$,
\begin{equation} \label{3.3}
		{\rm df}^{n+1}(0) = {\rm df}^n(0) \cdot \exp(-r^n\cdot h).
\end{equation}

Next, for a curve term $s_k$, $k =1,2,\cdots, K$, we calculate the zero rate from the model equation,
\begin{equation} \label{3.4}
		y^{n+1}_k = y^n_k + \Bigg[ \frac{f^n_k - r^n}{s_k} + \frac{\sigma^2(t^n,s_k)\, s_k}{2} \Bigg]\cdot h
		+  \langle \vec{\bf \sigma}(t^n,s_k) \cdot {\Delta}{\bf w}(t^n) \rangle,
\end{equation}
where the forward rate $f^n_k = f(t^n,s_k,h)$ is calculated from the known curve at $t^n$
\begin{equation} \label{3.5}
		f^n_k = \frac{1}{h}\, \ln \frac{{\rm df}^n(s_k)}{{\rm df}^n(s_k+h)}.
\end{equation}
When $y^{n+1}_{k}$ is obtained, we can calculate a predicted discount factor,
\begin{equation} \label{3.6}
		{\rm df}^{n+1}(s_k) = c\cdot {\rm df}^{n+1}(0) \cdot
			\exp\big(-y^{n+1}_k \cdot s_k \big) = {\rm df}^{n+1}(s_k,q),
\end{equation}
where we put back the Monte Carlo path index $q$. The constant $c$ in the equation is determined by a curve fitting across all paths  $q=1,2,\cdots, Q$ at the time step $t^{n+1}$  and  curve term $s_k$, i.e.,
\begin{equation} \label{3.7}
		\frac{1}{Q}\, \sum\limits_{q=1}^{Q} {\rm df}^{n+1}(s_k,q) = {\rm df}^0(t^{n+1}+s_k).
\end{equation}
By repeating the work above, we are able to finish the stochastic processes.

\section{\large Grid method}
\setcounter{equation}{0}

This model can also be implemented by a grid method. Like that in the BGM model, we take an approximation to the drift term in the dynamic equation. Here we take the one-factor model
(\ref{2.24}) as an example. We replace the drift term by a time-dependent function
$\theta_k(t)$,
\begin{equation} \label{4.1}
		\partial y_k = \theta_k(t) \, {\rm d}t + \sigma_k \, {\rm d}w, \hspace{5mm}
			k=1,2,\cdots,K.
\end{equation}
Thus, the stochastic processes $y_k(t)$ for $k = 1,2,\cdots,K$ become Markov variables, where   $\sigma_k = \sigma(t,s_k)$ still have the dual-term structure. The function $\theta_k(t)$   will be determined by the curve fitting, which is similar to the variable $c$ in (\ref{3.6}).

Grid methods are most successful in developing short rate models. There are many versions of grid methods. The suitable one to be extended to the zero rate model is the Krasker method, which has been used in the industry since the time of the Salomon Brothers. In the following, we first give a brief review to the Krasker method, because it cannot be found in the public literature. It starts with a grid probability field for a state variable $x$ on the stochastic differential equation
\begin{equation} \label{4.2}
		{\rm d}x = \sigma_0(t) \, {\rm d} w,
\end{equation}
where $\sigma_0(t) = \sigma(t,0)$ is the volatility of the short rate. It can be obtained from the far left-hand column in the model volatility matrix. We will build a tree-like grid field, such as the one in Figure~2, where half the space width at time-step $t$ is set to 4 times the standard deviation of $x(t)$, and the space width is then equally divided into a number of $I$ (e.g., $I = 40$ or 80) intervals $\Delta x^n$.

\input {./fig_401.tex}
\input {./fig_402.tex}

The transition probability from grid point $x^{n-1}_i$   to grid point $x^n_j$  is calculated in two steps, as shown in Figure~3. In the first step, a quaternary tree is applied to the starting point  $x^{n-1}_i$. Each tree branch is connected to an intermediate point  $x^n_*$,
\begin{equation} \label{4.3}
		x^n_* = x^{n-1}_i + \sigma_0(t^{n-1}) \cdot \Delta w,
\end{equation}
where $\Delta w$ and the probability distribution to reach  $x^n_*$ is
\begin{eqnarray}  \label{4.4}
	\Delta w &=& \Big\{ -\sqrt{3\Delta t},\, - \sqrt{\Delta t/3},\, \sqrt{\Delta t/3}, \,\sqrt{3\Delta t} \Big\},
		\nonumber \\
	p^{n-1}_{i*} &=&  \Big\{0.125,\, 0.375, \, 0.375, \, 0.125\Big\}.
\end{eqnarray}
In the second step, a grid point $x^n_j$  closest to $x^n_*$  is found. From a Taylor expansion on a variable $u$,
\begin{equation} \label{4.5}
		u_* = \frac{1}{2}\Big(\xi^2-\xi \Big)u_{j-1} + \Big(1-\xi^2\Big) u_j +
			\frac{1}{2}\Big(\xi^2+\xi \Big)u_{j+1},
\end{equation}
where  $\xi = (x^n_* - x^n_j)/\Delta x^n$. The coefficient in the formula is taken as the transition probability, i.e.,
\begin{equation} \label{4.6}
		p^n_{*,j-1} = \frac{1}{2}\Big(\xi^2-\xi \Big), \hspace{5mm}
			p^n_{*j} = \Big(1-\xi^2\Big), \hspace{5mm}
			p^n_{*,j+1} =  \frac{1}{2}\Big(\xi^2+\xi \Big).
\end{equation}
Thus, the transaction probability from the grid point $x^{n-1}_i$ to grid point $x^n_j$ is obtained by combining ({\ref{4.4}) and ({\ref{4.6}),
\begin{equation} \label{4.7}
		p^{n-1}_{ij} = \sum\limits_{*} p^{n-1}_{i*} \cdot p^n_{*j}.
\end{equation}

In the short rate model, the state variable $r$ is defined on each grid point $r^n_i = r(x^n_i)$  by a Gaussian normal mapping
\begin{equation} \label{4.8}
		r^n_i = x^n_i + \mu^n.
\end{equation}
The probability weighted discount factor at each grid point, i.e., Arrow-Debreu price $z^n_i$,  is obtained by a forward propagation
\begin{equation} \label{4.9}
		z^{n}_j = \sum\limits_{i=-I}^{I} z^{n-1}_i \cdot
				\exp \Big( -r^{n-1}_i\cdot \Delta t \Big) \cdot p^{n-1}_{ij},
\end{equation}
where at the tree root $z^0_0 = 1$. Thus, the $\mu^n$  is determined by the curve fitting condition at each time step:
\begin{equation} \label{4.10}
				\sum\limits_{j=-I}^{I} z^{n}_j = {\rm df}^0(t^{n}).
\end{equation}

Now we extend above approach to zero rates. In the grid method, the $K$ zero rates $y_k(t) = y(t,s_k)$, $k=1,2,\cdots,K$   are also defined on the grid point $x^n_i$,
\begin{equation} \label{4.11}
		(y_k)^n_i = a^n_k \cdot x^n_i + \mu^n_k,
\end{equation}
where $a^n_k$ is determined by 4 times the standard deviation of $y_k(x^n)$,
which can be calculated step by step from the cumulative variance using equation (\ref{4.1})
\begin{equation} \label{4.12}
		{\mathbb{V}} \big[y_k(x^n)\big] = {\mathbb{V}} \big[y_k(x^{n-1})\big]
			+ \sigma^2_k(t^{n-1})\cdot \Delta t^{n-1}, \hspace{5mm} n=1,2,\cdots,
\end{equation}
and the drift term $\mu^n_k$   will be determined by applying (\ref{2.6}) in the yield curve fitting, which becomes
\begin{equation} \label{4.13}
		\sum\limits_{i=-I}^{I}  \exp \Big\{ - (y_k)^n_i \cdot s_k  \Big\}  \cdot  z^n_i
			= {\rm df}^0(t^n+s_k).
\end{equation}
After that, a discount factor  at the grid point  $x^n_i$ is obtained by
\begin{equation} \label{4.14}
		{\rm df}^n_i(s_k) = z^n_i \cdot \exp \Big\{ - (y_k)^n_i \cdot s_k \Big\},
		\hspace{5mm} k = 1,2,\cdots,K,
\end{equation}
which serves as a yield curve ${\rm df}^n_i(s)$ for the pricing.

\section{\large Calculation of cash flows}
\setcounter{equation}{0}

Cash flow is another essential topic in pricing. Like any long rate model in which a yield curve is served as the state variable, there are two possible choices to perform the task. Suppose we stand at time $t^n$  of a Monte Carlo path. As shown in Figure~4,  we can calculate the forward cash flows by a discount factor curve at this time step,  ${\rm df}^n(s)$, to obtain cash flows, such as  $c^n_1$, $c^n_2$, $c^n_3$, $c^n_4, \cdots$. We can also do the work using other curves like ${\rm df}^{\alpha}(s)$, ${\rm df}^{\beta}(s)$, ${\rm df}^{\gamma}(s), \cdots$   on the following time steps, such as  $c^n_1$, $c^\alpha_1$, $c^\beta_1$, $c^\gamma_1, \cdots$, and discount the cash flows backward to time step $t^n$. From the view-point of trading practices, this is actually a question of what kind of instruments we use for hedging. A good example is a European swaption. Theoretically, its payoff depends on a underlying swap, which can be the function of either a long-term swap rate  $S(t)$, or a series of the short-term Libor rates $L_j(t)$
\begin{equation} \label{5.1}
	{\rm Swap}(t) = \big[ S(t) - K \big] \cdot
		\sum\limits_{j} P(t,T_j) \cdot \Delta T_j
		= \sum\limits_{j} \big[ L_j(t) - K \big] \cdot  P(t,T_j) \cdot \Delta T_j.
\end{equation}
However, in trading practices, traders always use long-term swap to hedge swaption instead of using short-term Libors, because they are too expensive.
\input {./fig_501.tex}

In the zero rate model, we divide forward cash flows into two groups. One group is called the vanilla cash flow, in which there is no caplet feature (such as $\max(F - K, 0)$) in the payoff formula. In this case, we directly use the discount factor curve  ${\rm df}^n(s)$ on the current time step $t^n$  to calculate the cash flow. For example, for a swap starting at time  $t^n$ that has $m$ periods of payments on the payment dates  $t^n+S_1, t^n+S_2, \cdots, t^n+S_m$, its value is calculated by
\begin{equation}  \label{5.2}
		{\rm SwapVal} = {\rm df}^n(0) - {\rm df}^n(S_m) -
			 X \cdot \sum\limits_{j=1}^{m} {\rm df}^n(S_j)\cdot \Delta S_j.
\end{equation}

The other group is the non-vanilla cash flow. A cap is an example of a non-vanilla cash flow, in which different caplets $\max(F_n - K, 0)$, $\max(F_\alpha - K, 0)$, $\max(F_\beta - K, 0), \cdots$  are calculated by a different curve ${\rm df}^n(s)$, ${\rm df}^\alpha(s)$, ${\rm df}^\beta(s), \cdots$. Non-vanilla cash flows are often seen in structured products, e.g., a CMS spread option with payoff
$\max({\rm CMS30Y}-{\rm CMS2Y}, 0)$. In the same way, the indexes CMS30Y and CMS2Y are calculated using the (\ref{5.2}) with one curve, either ${\rm df}^n(s)$,  or ${\rm df}^\alpha(s)$, ${\rm df}^\beta(s)$,
${\rm df}^\gamma(s), \cdots$, and the option max is performed in the term root $s=0$ of each curve.

\section{\large Global volatility calibration}
\setcounter{equation}{0}

A dual-term structure model has two terms in the model volatility. In the numerical computation, the model volatility parameters $\{\sigma_{ik}\}$  form a two-dimensional array. If the numbers of model parameters are bigger than the number of market instruments, an exact model calibration becomes possible. However, an advanced numerical method is still needed to do the calibration.

\input {./fig_601.tex}

We use the Lagrange multiplier method to calibrate the model. First, the definition region of model volatility is divided into many rectangle elements, $A, B, C, D$, etc., as in Figure~5. The model volatility surface is discretized into a piecewise linear function on each element,
\begin{equation}  \label{6.1}
		\sigma(t,s) = \sigma_A(t,s)\cdot I_A(t,s) +  \sigma_B(t,s)\cdot I_B(t,s) + \cdots,
\end{equation}
where $I_A(t,s)$, $I_B(t,s), \cdots$  are indication functions such that
\begin{equation}  \label{6.2}
		I_A(t,s) = \left\{ \begin{array}{ll}
					1, & (t,s) \in A,  \\[2mm]
					0, & {\rm otherwise}.  \end{array} \right.
\end{equation}
And $\sigma_A(t,s)$, $\sigma_B(t,s), \cdots$ are bi-linear functions defined on the rectangle elements, e.g.
\begin{equation}  \label{6.3}
		\sigma_A(t,s) = \sigma_1\cdot(1-\xi)(1-\eta)+ \sigma_2\cdot \xi(1-\eta) +
			\sigma_4\cdot (1-\xi)\eta + \sigma_3\cdot \xi\eta,
\end{equation}
where, $\xi = (t-t^n) /\tau$, $\eta = (s-s_j) /\delta$.

We look for a smooth model volatility surface $\sigma(t,s)$. The smoothness of the surface is characterized by its area, which, after ignoring a constant, can be approximately measured by
\begin{equation}   \label{6.4}
		S = \sum\limits_{\omega =A,B,\cdots} \frac{3}{2} \int\!\!\!\int\limits_{\!\!\!\!\omega}
				\bigg[ 	\bigg(\frac{\partial \sigma_\omega}{\partial t} \bigg)^2 +
						\bigg(\frac{\partial \sigma_\omega}{\partial s} \bigg)^2 \,
				\bigg] {\rm d}t {\rm d}s
\end{equation}
where the coefficient 3/2 is chosen for deriving a uniform formula below. The smoother the surface is, the smaller the area will be. Thus, the solution $\sigma(t,s)$  should make $S$ reach a minimum.

The market conditions are expressed by the prices of swaption or cap/floor instruments. Assume there are $M$ market instruments. The model prices calculated from the model volatility are denoted by
$V_1,V_2,\cdots,V_M$, while the market quotes are $\hat{V}_1,\hat{V}_2,\cdots,\hat{V}_M$. Thus the relative error for $m$-th instruments is
\begin{equation}    \label{6.5}
		E_m = \frac{V_m}{\hat{V}_m} - 1.
\end{equation}
Apparently, $E_m$  is a function of  $\sigma(t,s)$, i.e., $E_m = E_m(\sigma_1,\sigma_2,\cdots,\sigma_K)$. The total error can be expressed by an $L2$ norm
\begin{equation}    \label{6.6}
		||E||^2 = \frac{1}{M} \sum\limits_{m=1}^{M} E_m^2.
\end{equation}
Therefore, the formulation for model calibration by the Lagrange method is: Finding a model volatility surface  $\sigma=\sigma(t,s)$, which ensures $||E||=0$  and $S$ reach the minimum.

To solve this minimum problem, we introduce a set of Lagrange multipliers
$\lambda_1$, $\lambda_2,\cdots$, $\lambda_M$  and construct an objective function
\begin{equation}    \label{6.7}
		L = \sum\limits_{\omega =A,B,\cdots} \frac{3}{2} \int\!\!\!\int\limits_{\!\!\!\!\omega}
				\bigg[ 	\bigg(\frac{\partial \sigma_\omega}{\partial t} \bigg)^2 +
						\bigg(\frac{\partial \sigma_\omega}{\partial s} \bigg)^2 \,
				\bigg] {\rm d}t {\rm d}s
			+ \sum\limits_{m=1}^M \lambda_m\,E_m(\sigma_1,\sigma_2,\cdots,\sigma_K).
\end{equation}
We take partial derivatives of $L$ with respect to  $\sigma_k$ to obtain $K$ algebraic equations, e.g. for $\sigma_1$  we have		
\begin{eqnarray}    \label{6.8}
		\frac{\partial L}{\partial \sigma_1}
			= (\tau + \delta)\sigma_1 + \Big( \frac{\tau}{2} - \delta \Big)\sigma_2 +
			 \Big( \frac{\delta}{2} - \tau \Big)\sigma_4 -\frac{\tau+\delta}{2} \sigma_3 \hspace{1cm}
			\nonumber\\
		+ [B] + [C] + [D] + \sum\limits_{m=1}^{M} \lambda_m \frac{\partial E_m}{\partial \sigma_1} = 0,
\end{eqnarray}
where only the part of the scheme in rectangle $A$ is listed; other parts are represented by the symbols $[B]$, $[C]$ and $[D]$, which can be easily obtained by a mirror reflection from the scheme in rectangle $A$. In numerical computation, knowing the scheme in rectangle $A$ is enough for programming, and the derivative $\partial E_m/\partial \sigma_k$  is calculated numerically. If $\sigma_k$  is located at a corner point or a boundary point, the corresponding equation can be obtained from
(\ref{6.8}) by removing the related scheme parts from the non-existing rectangles.

The unknown in (\ref{6.8}) include all $\sigma_1$, $\sigma_2,\cdots$, $\sigma_K$  and
$\lambda_1$, $\lambda_2, \cdots$, $\lambda_M$. Thus, we still need $M$ equations to close the system. We then make an approximation on  $E_m = E_m(\sigma_1$, $\sigma_2,\cdots$, $\sigma_K)$ at an initial point  $E_m(\dot{\sigma}_1, \dot{\sigma}_2,\cdots, \dot{\sigma}_K)$ by a Taylor expansion,
\begin{equation} \label{6.9}
		E_m(\sigma_1, \cdots, \sigma_K) = E_m(\dot{\sigma}_1, \cdots, \dot{\sigma}_K)
				+ \frac{\partial E_m}{\partial \dot{\sigma}_1} {\rm d}\sigma_1 + \cdots
				+ \frac{\partial E_m}{\partial \dot{\sigma}_K} {\rm d}\sigma_K.
\end{equation}
Assuming $(\sigma_1, \sigma_2,\cdots, \sigma_K)$  is a solution point that makes  $E_m(\sigma_1, \sigma_2,\cdots, \sigma_K)=0$ , we then obtain other $K$ algebraic equations
\begin{eqnarray} \label{6.10}
			\frac{\partial E_m}{\partial \dot{\sigma}_1} \sigma_1 +
			\frac{\partial E_m}{\partial \dot{\sigma}_2} \sigma_2 + \cdots
				+ \frac{\partial E_m}{\partial \dot{\sigma}_K} \sigma_K \hspace{5cm}
				\nonumber \\[2mm] =
			\frac{\partial E_m}{\partial \dot{\sigma}_1} \dot{\sigma_1} +
			\frac{\partial E_m}{\partial \dot{\sigma}_2} \dot{\sigma_2} + \cdots
				+ \frac{\partial E_m}{\partial \dot{\sigma}_K} \dot{\sigma_K}
				-E_m(\dot{\sigma_1},\cdots, \dot{\sigma_K}),
				 \nonumber\\[2mm]
			m = 1,2,\cdots,M. \hspace{1cm}
\end{eqnarray}
Thus there are $K + M$ equations in (\ref{6.8}) and (\ref{6.10}). The number of known variables is also $K + M$, which may be solved simultaneously. Since the equation is non-linear, the first solution   will be approximated. Therefore, an iteration is needed. There is a big challenge in solving the system of (\ref{6.8}) and (\ref{6.10}) numerically, since the computation time and computer memory will increase tremendously when the number of volatility parameters   increase. The parallel computing technique is needed to do the work.  However, the iteration converges very fast. An example will be presented in the Section 8.

\section{\large Bucket vega}
\setcounter{equation}{0}

The bucket vega is a change of a deal's present value (PV) caused by a 1\% change in a market volatility component. Suppose there are $M$ market volatilities. Thus, the bucket vegas will also consist of $M$ components. To calculate a bucket vega, we need a mapping from the model volatility space to the market volatility space. Recall that the volatility has a dual-term structure; thus, the mapping should be a two-dimensional mapping. One may consider using the same number of model parameters as $M$, to make a one-to-one mapping. Our work showed that this would result in an un-smooth model volatility surface, which would make calculating bucket vegas impossible. In the zero rate model, the number of model parameters $K$ is larger than $M$. Following is a method to calculate the bucket vegas.

Let $U$ be the PV of a deal, which is a function of model volatility components  $(\sigma_1,\sigma_2,\cdots,\sigma_K)$. Thus, the incremental of $U$ can be represented by
\begin{equation} \label{7.1}
		{\rm d} U = \big({\rm d}\sigma_1, {\rm d}\sigma_2, \cdots, {\rm d}\sigma_K \big) \cdot {\bf b},
\end{equation}
where ${\bf b} = (\partial U / \partial \sigma_k )$  is a vector of model vega components, which can be calculated numerically. It is important to assume all components of {\bf b} are non-zero, otherwise the corresponding component $\sigma_k$  is un-related and can be removed from the computation. Let $(v_1,v_2,\cdots,v_M)$  be the market volatility for each swaption instrument, which we assume to be lognormal volatilities in this paper. Let also  $(V_1,V_2,\cdots,V_M)$  be the PVs of each instrument. The market vegas of these $M$ instruments are represented by
\begin{equation}  \label{7.2}
		g_m = \frac{{\rm d}V_m}{{\rm d} v_m}, \hspace{1cm} m=1,2,\cdots,M,
\end{equation}
which can be written in a vector form as
\begin{equation}  \label{7.3}
		({\rm d}V_1, {\rm d}V_2,\cdots, {\rm d}V_M) = ({\rm d}v_1, {\rm d}v_2,\cdots, {\rm d}v_M)
			\cdot {\bf G},
\end{equation}
where ${\bf G} = {\rm diag}(g_1,g_2,\cdots,g_M)$  is a diagonal matrix. The changes of PVs of $M$ instruments can also be expressed by the changes of model volatilities as
\begin{equation}  \label{7.4}
		({\rm d}V_1, {\rm d}V_2,\cdots, {\rm d}V_M) =
			({\rm d}\sigma_1, {\rm d}\sigma_2,\cdots, {\rm d}\sigma_K) 	\cdot {\bf A},
\end{equation}
where ${\bf A} =(\partial V_m /\partial \sigma_k)$ is a Jacobean matrix, and also can be calculated numerically. Thus, we have
\begin{equation}  \label{7.5}
		({\rm d}v_1, {\rm d}v_2,\cdots, {\rm d}v_M) =
			({\rm d}\sigma_1, {\rm d}\sigma_2,\cdots, {\rm d}\sigma_K) 	\cdot
				{\bf A} \cdot {\bf G}^{-1}.
\end{equation}
Now we look for the bucket vegas of the deal,  ${\bf x} = (x_1,x_2,\cdots,x_M)^T$, which could be expressed by
\begin{equation}   \label{7.6}
		{\rm d}U = ({\rm d}v_1, {\rm d}v_2,\cdots, {\rm d}v_M)\cdot {\bf x}
			 =  ({\rm d}\sigma_1, {\rm d}\sigma_2,\cdots, {\rm d}\sigma_K) 	\cdot
				{\bf A} \cdot {\bf G}^{-1} \cdot {\bf x}.
\end{equation}
Comparing (\ref{7.1}) and (\ref{7.6}), we have equations for solving the bucket vega {\bf x}:
\begin{equation}   \label{7.7}
		{\bf A} \cdot {\bf G}^{-1} \cdot {\bf x} = {\bf b}.
\end{equation}
Since $K>M$,  {\bf A} is a high matrix. Thus, the equation (\ref{7.7}) can be solved by the Singular-Value-Decomposition (SVD) method [13].

\section{\large Pricing examples}
\setcounter{equation}{0}

We are going to present five pricing examples by the zero rate model in this section. The valuation date is set to 2015-08-03. The market condition used in the pricing was obtained from Reuters on the valuation date, which consists of USD LIBOR swap rates, the USD over-night swap (OIS) rate (for building discount curve), and the USD LIBOR swaption log-normal type volatility matrix. Except for the 3-month LIBOR rate, which is 0.31\% and the 6-month LIBOR rate, which is 0.49\%, all other swap rates and volatilities are shown in the Table~1.

\input {./table_801.tex}
\input {./fig_801.tex}
\input {./fig_802.tex}

The first example is the global calibration to the swaption matrix. We took $21\times 21=441$ model volatility parameters  $\sigma_k$ for calibrating $9\times 10$ swaption instruments in Table~1. We started with an equal value $\sigma_k = 0.01$ for all parameters, where $||E^{(0)}|| = 9.6128$. Then we only took three loops to reach a satisfactory result:  $||E^{(1)}|| = 0.7081$,  $||E^{(2)}|| = 0.0147$, and  $||E^{(3)}|| = 0.0004$. The convergence is very fast. In the following examples, the similar calibrations  will be done.

The second example is the pricing on conventional deals. We consider Bermudan options to enter or cancel a 10-year break-even swap. The swap with notional \$10,000 USD starts on 2015-08-05 for 10 years, pays a fixed rate of 2.281\% (10-year break-even rate) semi-annually with a day-count of 30/360, and receives 3-month USD LIBOR quarterly with a day-count of ACT/360. The option starts in 6 months and exercises semi-annually.

The 1-factor and 3-factor zero rate model proposed in this paper have been applied to compute the option values of the two deals. In the cases of using the Monte Carlo method, 10,000 paths have been used in the simulation. The option values of the deals were calculated by the Longstaff-Schwartz's Least-squared approach [8], which have been separated into 19 exercise-periods during the price rolling-back process, and are shown in Figures~6 and 7, respectively. The results were compared to those from a 1-factor short rate model (BDT or Hull-White models), and a 3-factor short rate model (2-Plus model).

From Figures~6 and 7, we can see that for conventional deals, the zero rate model gives very close results with those of the traditional short rate models. The smoothness of the call option distribution is not as good as that from the grid method, which we believe was from the Least-squared approach in calculating the Bermudan option. However, from examples below, we will see that this is only a minor drawback. Although the short rate models give smooth results for simple structured deals, for complicated structures, the short rate model would give incorrect results, or even no solution.

The third example is a 10-year callable range-accrual inverse floater deal. The swap starts on 2015-8-19, the notional is \$10,000 USD. In the deal we pay a coupon $\max(6\%-{\tt LIBOR\_6M},0)\cdot n/N$  semi-annually with a day-count of 30/360, where $n$ is the number of days in a payment period in which ${\tt CMS\_30Y}-{\tt CMS\_2Y} >0$, $N$ is the total number of days of the payment period, and we receive
${\tt LIBOR\_6M}$   semi-annually with a day-count of ACT/360. We have a call option which starts in 6 months and exercises semi-annually.

\input {./fig_803.tex}

Figure~8 shows the distribution of the swap value and option value along the exercise periods, which have been calculated by different models listed in the graph. As we can see from the graph, the swap value obtained from a 1-factor zero rate model of this paper agrees well with the short rate models of 1-factor BDT/HW and 3-factor 2-Plus. The 3-factor zero rate model of this paper gives more freedom to the changes of CMS rates, thus leads to a high swap value. A remarkable feature of the option value distribution is shown in the graph. While the two short rate models basically give a flat distribution for the option value, the zero rate model of this paper reveals that the option values rise much higher in the short end (from exercise period 1 to 6). A straight-forward analysis of the deal will show that the result of the zero rate model is more reasonable. Because the current 6-month LIBOR rate is just about 0.4\%, we pay much higher coupon (almost 5.6\%) than that we receive on the front end. This means that the swap is deep out-of-money in the periods from 1 to 6, thus the call option should be deep in-the-money.

\input {./table_802.tex}
\input {./table_803.tex}

The fourth example calculates bucket vegas on a Bermudan swaption. As we mentioned earlier in this paper, bucket vegas can be calculated in the short rate model by specifically picking up a series of volatility instruments along an anti-diagonal line. However, this volatility pick-up will involve a human factor in the pricing. Since the zero rate model introduced in this paper can make a global calibration to the volatility matrix, it is possible to let the model choose volatility to calculate the bucket vegas. We consider two Bermudan options that both exercise annually, one is to enter a 10-year swap, the other is to cancel a 10-year swap. For a cleaner picture on the bucket vegas, we calibrate the model to a $10\times 10$ years' swaption matrix. The method in Section~7 has been used to calculate the bucket vegas. Due to the un-smooth problem in the Monte-Carlo method for calculating the Bermudan options, we use the grid method to do the work.

\input {./table_804.tex}
\input {./table_805.tex}

Tables~2 and 3 are the distributions of bucket vegas on the Bermudan swaption. The results are promising, since we already have a triangular distribution. We also see the biggest vega components appearing in the anti-diagonal lines. Due to the finite-difference approximation in the numerical method, there are still some oscillations in the vega distribution.

The fifth example is for the calculation of bucket vegas on the structured products. In hedging and risk management, computing bucket vegas of structured products has always been one of the most important but difficult tasks. The zero rate model proposed in this paper contributes a promising step in the work. Let us consider a CMS spread swap with notional \$1,000,000 USD, which starts on 2018-08-06 (three years from now) and lasts for 2 years, pays a fixed-rate of 1\% and receives  $2\times \max({\tt CMS\_10Y} - {\tt CMS\_2Y}, 0)$. Both legs pay quarterly with a day-count of ACT/360. First, we calculated bucket vegas for this structured-swap, shown in Table~4. The figure shows strong components on the 4Y$\times$2Y  and 4Y$\times$10Y  terms, corresponding to the CMS indexes in the deal.

Then, we added a cancelation option on the swap start date. The result (sum of the swap and option vegas) has been shown in Table~5. In trading practice, traders have been puzzled for long time by how to hedge callable structured deals like this. The zero rate model gives an interesting answer to the problem. Although the cancel option changes the vega values, the strong vega components are still located around the  4Y$\times$2Y  and 4Y$\times$10Y  term points. Thus, traders can use the same underlying CMS instruments to hedge both the no-call swap and the callable swap deals.

\section{\large Conclusions}
\setcounter{equation}{0}

In this paper, we have proposed a deal-term structured model for pricing interest rate options and structured products. The underlying state variables used by the model are a series of long-term zero rates of zero-coupon bonds at different maturities, which are driven directly by one or more Brownian motions. The model volatility has two terms, which takes a matrix format that makes an exact model calibration to the entire swaption matrix possible. Then, we have presented numerical schemes for using the model in this paper, both Monte Carlo and grid methods. We have also derived the Lagrange multiplier formulas for model calibration, and the Jacobian inverse equations for calculating bucket vegas. At the end, we provided five numerical examples that show the model's abilities to solve the most difficult problems in financial modeling.

\section{\large Appendix}
\setcounter{equation}{0}

We will derive the dynamic equation for the zero rate model in a simple market environment in this appendix, which can be compared to the equation in the continuous market. The simple market has a direct background in a numerical modeling, in which the continuous time $t$ is divided into a set of discrete steps and the interest rate is treated as right continuous and valid for the step period.

First, we define a simple market. Let $(\Omega, {\cal F}, ({\cal F}_t)_{t\geq0}, \mathbb{P})$ be a complete probability space, where $({\cal F}_t)_{t\geq0}$ is a filtration satisfying usual conditions.  A simple market is an idealized scenario of a real market, in which any interest rate, say $(r_t)_{t\geq0}$, is right continuous and ${\cal F}_{t}$-adapted. We call such an interest rate process as simple process.

Without a repeat work, we continue to use  several financial quantities from the simple market environment, such as
zero coupon bond $P(t,T)$, zero coupon rate $y(t,T-t)$, bank account $B(t)$, short rate $r(t)$, and forward rate $f(t,T-t,h)$. We shall assume that for any $h\geq0$,
\begin{eqnarray} \label{2.5}
		{\mathbb{E}}\Big[P(t+h, T+h)\Big] <\infty,\quad
		{\mathbb{E}}\Big[\exp\Big\{|f(t, T-t, h)| h\Big\} \Big]<\infty.
\end{eqnarray}
And, when $h\to 0$, the instantaneous forward rate $f(t,T-t,0)$ exist.

Then we come to the fundamental pricing theory. It states that in a non-arbitrage market, there exists a risk-neutral measure, with the bank account as numeraire, so that the ratio
\begin{eqnarray} \label{2.6}
		\frac{P(t, T)}{B(t)}=\exp\Big\{-y(t, T-t)\cdot (T-t)-\int_0^tr(\tau)d\tau\Big\}
\end{eqnarray}
is an ${\cal F}_t$-martingale for time period $0\leq t\leq T$.
Specifically, for a bond matured on $\hat{T}$, it is equivalent to that the conditional expectation at time $t$
\begin{eqnarray} \label{2.7}
		{\mathbb E}\Bigg[ \frac{P(t+h, \hat{T})}{B(t+h)}\bigg{|}{\cal F}_t\Bigg] =	
			\frac{P(t, \hat{T})}{B(t)}
\end{eqnarray}
holds. Our model for the zero rate $y(t, T-t)$  will be built up based on the martingale condition of (\ref{2.7}) with $\hat{T} = T+h$. The model is stated by following Theorem.

{\bf Theorem:} Let $y(t,T-t)$, $r(t)$ and $f(t,T-t,0)$ be zero rate, short rate and forward rate of a zero-coupon bond with life $s=T-t$ in the simple market, and satisfy the conditions of (\ref{2.5}). Let $\partial y(t, s) = y(t+ {\rm d}t, s) - y(t, s)$ be the partial derivative of zero rate with respect to the first argument $t$,  which follows a normal stochastic process
\begin{eqnarray} \label{2.8}
			\partial y(t, s) \,= \, \mu(t, s){\rm d}t+\sigma(t, s) \,{\rm d} w_t,
\end{eqnarray}
where, $\sigma(t, s)$ is a deterministic volatility function, $w_t=w(t)$ is ${\cal F}_t$-Brownian motion and $\mu(t, s)$ is an ${\cal F}_t$-adapted drift process. Then, under the non-arbitrage condition, the drift function
has the following form:
\begin{eqnarray} \label{2.9}
		\mu(t,s) = \frac{ f(t,s,0) - r(t) } {s} + \frac{1}{2} \sigma^2(t,s)\, s.
\end{eqnarray}

Proof: We consider a zero-coupon bond $P(t, T+h)$  that matures on $T+h$. Using the forward rate $f(t, T-t, h)$, (\ref{2.2}) and (\ref{2.4}) are combined as
\begin{eqnarray} \label{2.10}
		\frac{P(t, T+h)}{B(t)}
		=\exp\Big\{-y(t, T-t)\cdot (T-t)- f(t,T-t,h)\cdot h-\int_0^t r(\tau)d\tau\Big\}.
\end{eqnarray}

We further consider the price change of bond $P(t, T+h)$  in one time-step from  time $t$ to $t + h$. And at time $t + h$, its zero rate is denoted by $y(t+h, T-t)$.
Thus, applying (\ref{2.6}) to the bond at $t+h$, we have,
\begin{eqnarray} \label{2.11}
		\frac{P(t+h, T+h)}{B(t+h)}
		=\exp\Big\{-y(t+h, T-t)\cdot (T-t)-\int_0^{t+h}r(\tau)d\tau\Big\}.
\end{eqnarray}

From (\ref{2.8}), we express the model equation as
\begin{eqnarray} \label{2.12}
		y(t+h, s) \,= \, y(t,s) + \int\limits_t^{t+h} \mu(\tau, s){\rm d}\tau
					+\int\limits_t^{t+h} \sigma(\tau, s) {\rm d} w_\tau.
\end{eqnarray}
Substituting (\ref{2.12}) into (\ref{2.11}), the left hand side of (\ref{2.7}) is equal to
\begin{eqnarray} \label{2.13}
		{\mathbb E}\Bigg[ \exp\bigg{\{} -\Big(y(t, s)+\int\limits_t^{t+h}\mu(\tau, s){\rm d}\tau
			+\int\limits_t^{t+h}\sigma(\tau, s){\rm d}w_\tau\Big)\cdot s -\int\limits_0^{t+h}r(\tau){\rm d}\tau \bigg{\}}
			 \bigg| {\cal F}_t
							\Bigg]  \nonumber \\
		 = \exp\bigg\{ - y(t,s)\, s -\int\limits_0^{t}r(\tau){\rm d}\tau \bigg\}
						\cdot \, {\mathbb E}\bigg[ e^{ X(t+h) -X(t)}  \bigg| {\cal F}_t  \bigg], \hspace{1cm}
\end{eqnarray}
where
\[
		X(t) =  - s\! \int\limits_0^{t}\mu(\tau,s){\rm d}\tau
							-s \!  \int\limits_0^{t} \sigma(\tau,s) {\rm d}w_\tau
															- \int\limits_0^{t} r(\tau){\rm d}\tau.
\]
The right hand side of (\ref{2.7}) is (\ref{2.10}). Thus, (\ref{2.7}) becomes
\begin{eqnarray} \label{2.14}
				{\mathbb E}\bigg[ e^{ X(t+h) -X(t)}  \bigg| {\cal F}_t  \bigg] = e^{-f(t,s,h) \cdot h},
\end{eqnarray}
which implies that $f(t,s,h)$ is ${\cal F}_t$ measurable and hence $f(t,s,0) := \lim_{h\downarrow0}f(t, s, h)$ is also  ${\cal F}_t$ measurable. By Ito's formula, we obtain
\begin{eqnarray} \label{2.14a}
		e^{X(t)} -e^{X(0)} &=& \int\limits_0^t e^{X(\tau)} {\rm d}X_\tau
			+ \frac{1}{2} \int\limits_0^t e^{X(\tau)} {\rm d}\langle X_\tau \rangle \hspace{2cm}
				\nonumber \\[1mm]
		 &=& \int\limits_0^{t}  e^{X(\tau)} \bigg\{ -s \mu(\tau,s) -  r(\tau) + \frac{1}{2}s^2\sigma^2(\tau,s)
							 \bigg\}{\rm d}\tau
							- s \!  \int\limits_0^{t} e^{X(\tau)} \sigma(\tau,s) {\rm d}w_\tau. 	
 \nonumber \\& &\hspace{5mm}		
\end{eqnarray}
We make the same calculation for  $e^{X(t+h)}$ and then substract it from (\ref{2.14a}),
\begin{eqnarray} \label{2.14b}
		\; e^{X(t+h)} =  e^{X(t)} &+&
			 \int\limits_t^{t+h}  e^{X(\tau)} \bigg\{\! -s \mu(\tau,s) -  r(\tau) + \frac{1}{2}s^2\sigma^2(\tau,s)
							 \bigg\}{\rm d}\tau
        \nonumber \\[1mm]
							&-& s \!  \int\limits_t^{t+h} e^{X(\tau)} \sigma(\tau,s) {\rm d}w_\tau.
\end{eqnarray}
Thus, we obtain
\begin{eqnarray} \label{2.14c}
		&& {\mathbb E}\bigg[ e^{ X(t+h) -X(t)}  \bigg| {\cal F}_t  \bigg]
        \nonumber \\[2mm]	
					 &=&
				{\mathbb E}\Bigg[ 1+ e^{-X(t)} \!\!
			 \int\limits_t^{t+h}  e^{X(\tau)} \bigg\{\! -s \mu(\tau,s) -  r(\tau) + \frac{1}{2}s^2\sigma^2(\tau,s)
							 \bigg\}{\rm d}\tau
									 \bigg| {\cal F}_t  \Bigg].
\end{eqnarray}
And (\ref{2.14}) implies for $h > 0$,
\begin{equation}
	{\mathbb E}\bigg[ \; \xi_h \;\bigg| {\cal F}_t  \bigg] = 0,
\end{equation}
where
\[
	\xi_h = \frac{1}{h} \Bigg\{ e^{f(t,s,h)h} \Bigg(
					1+ e^{-X(t)} \!\!
			 \int\limits_t^{t+h}  e^{X(\tau)} \bigg\{\! -s \mu(\tau,s) -  r(\tau) + \frac{1}{2}s^2\sigma^2(\tau,s)
							 \bigg\}{\rm d}\tau \Bigg) - 1 \Bigg\}.
\]
In the simple market where $\mu(t,\cdot)$, $r(t)$  and  $\sigma(t,\cdot)$ are right continuous, then
\begin{eqnarray} \label{2.16}
	\lim\limits_{h\downarrow 0} \xi_h \;\stackrel{a.s.}{=}\; - \mu(t,s) s
				 + f(t,s,0) - r(t) + \frac{1}{2}\sigma^2(t,s) s^2 \; =: \; \xi_0.
\end{eqnarray}
We shall prove that $\xi_0 = 0$ (a.s.).

To do this, we define
\begin{eqnarray} \label{2.17}
			A_{+} = \{\xi_0 > 0\} \in {\cal F}_t, \hspace{5mm} A_{-} = \{\xi_0 < 0\} \in {\cal F}_t.
\end{eqnarray}
We need to prove ${\mathbb P}[A_+] = 0$ and ${\mathbb P}[A_-] = 0$.
Because of $A_{+} \in {\cal F}_t$, by a property in conditional probability,
\begin{eqnarray} \label{2.19}
		 \int_{A_+} \xi_h {\rm d} {\mathbb P}
			= \int_{A_+} {\mathbb E}\bigg[\xi_h  \Big| {\cal F}_t \bigg] \, {\rm d} {\mathbb P}
			= \int_{A_+} 0\, {\rm d} {\mathbb P} = 0.
\end{eqnarray}
Thus, let $h\to 0$, we have
\begin{eqnarray} \label{2.21}
		 {\mathbb E}\bigg[\xi_0 \cdot I_{A_+} \bigg]
			= \lim\limits_{h\downarrow 0} {\mathbb E}\bigg[ \xi_h \cdot I_{A_+} \bigg]
			=  \lim\limits_{h\downarrow 0} \int_{A_+} \xi_h \, {\rm d} {\mathbb P} = 0.
\end{eqnarray}
If ${\mathbb P}[A_+] > 0$, we can define $A_n = \{\xi_0 > 1/n \}$. Thus, $A_+ = \cup_{n\ge 1} A_n $,
and there exists $n \ge 1$ such that ${\mathbb P}[A_n] > 0$. Then,
\begin{eqnarray} \label{2.22}
		 {\mathbb E}\bigg[\xi_0 \cdot I_{A_+} \bigg]
			= \int_{A_+}\xi_0\, {\rm d} {\mathbb P}
			= \int_{A_n}\xi_0\, {\rm d} {\mathbb P}  +  \int_{A_+\backslash A_n}\xi_0\, {\rm d} {\mathbb P}.
\end{eqnarray}
Since we have $\xi_0 > 1/n$ on $A_n$, and $\xi_0 > 0$ on $A_+\backslash A_n$,
\begin{eqnarray} \label{2.23}
		 {\mathbb E}\bigg[\xi_0 \cdot I_{A_+} \bigg]
			\ge  \int_{A_n} \frac{1}{n} \, {\rm d} {\mathbb P}  +  0 = \frac{1}{n} {\mathbb P}\big[ A_n \big] > 0.
\end{eqnarray}
This contradicts to (\ref{2.21}), which implies that ${\mathbb P}[A_+] = 0$. Similar reason also gives ${\mathbb P}[A_-] = 0$.
Thus, ${\mathbb P}[\{\xi_0 = 0\}] = 1$, we obtain the (\ref{2.9}) and complete the proof.

\vspace{4mm}

{\bf Acknowledgement}: The author would like to thank his former managers, Ranjit Bhattacharjee and Donald Chin at the Derivatives Research Group of Citigroup Global Markets, for proposing this project to him in 2006. The author would also thank Professor HE Hui at Beijing Normal University for the help to prove the Throrem in the paper.  The uses of China's Tianhe-2 supercomputer, located at the National Supercomputer Center in Guangzhou, for calculating the examples in the paper are also greatly acknowledged.

\section*{\large References}

\begin{enumerate}

\item[1.] Vasicek O. (1977): An equilibrium characterization of the term structure, {\it J. Financial Economics}, {\bf 5}: 177-188.


\item[2.]	Cox, J.C., J.E. Ingersoll and S.A. Ross (1985): A Theory of the Term Structure of Interest Rates. {\it Econometrica}, {\bf 53}: 385-407.

\item[3.]	Ho, T.S.Y., S.B. Lee (1986): Term structure movements and pricing interest rate contingent claims, {\it Journal of Finance}, {\bf 41}.

\item[4.]	Hull, J. and A. White (1990): Pricing interest-rate derivative securities, {\it The Review of Financial Studies}, {\bf 3} (4): 573-592.

\item[5.]	Black, F., E. Derman and W. Toy (January-February 1990): A One-Factor Model of Interest Rates and Its Application to Treasury Bond Options. {\it Financial Analysts Journal}: 24-32.

\item[6.]	Black, F. and P. Karasinski (July-August 1991): Bond and Option pricing when Short rates are Lognormal. {\it Financial Analysts Journal}: 52-59.

\item[7.]	Longstaff, F.A. and E.S. Schwartz (1992): Interest Rate Volatility and the Term Structure: A Two-Factor General Equilibrium Model. {\it Journal of Finance}, {\bf 47} (4): 1259-82.

\item[8.]	Longstaff, F.A. and E.S. Schwartz (2001):  Valuing American Options by Simulation: A Simple Least-Squares Approach. {\it The Review of Financial Studies Spring}, {\bf 14} (1): 113-147.

\item [9.]	Heath, D., R. Jarrow, and A. Morton (1990): Bond Pricing and the Term Structure of Interest Rates: A Discrete Time Approximation. {\it Journal of Financial and Quantitative Analysis}, {\bf 25}: 419-440.

\item [10.]	Brace, A., D. Gatarek, and M. Musiela (1997): The Market Model of Interest Rate Dynamics, {\it Mathematical Finance}, {\bf 7} (2): 127-154.

\item [11.]	Johnson S. (2006): Numerical Methods for the Markov Functional model, {\it Wilmott Magazine}, Jan: 68-75.

\item[12.]	Hunt, P.J., and J.E. Kennedy and A. Pelsser (2000): Markov-Functional Interest Rate Models, {\it Finance and Stochastics},
{\bf 4}: 391-408.

\item[13.]	Press, W.H., S.A. Teulolsky, W.T. Vetterling and B.P. Flannery (1992): {\it Numerical Recipes in C}. Cambridge University Press, 59-70.

\end{enumerate}

\end{document}

%% file: fig_301.tex
  

 


\begin{figure}[hbt] 

\[  \beginpicture   \setlinear 

\setcoordinatesystem units <10mm,10mm>
\setplotarea x from -1 to 10, y from 0 to 7

\thinlines

\arrow <1,5mm>   [0.25,0.75] from  1.1 0.0 to 10.0 0.0
\arrow <1,5mm>   [0.25,0.75] from  0.0 0.0 to  0.0 7.0

\plot  0    0  1 0  /
\plot  0.9  -0.1  1.1 0.1  /
\plot  1.0  -0.1  1.2 0.1  /

\plot  3.0  0.0  9.0 6.0  /
\plot  3.5  0.5  3.5 6.3  /
\plot  4.5  1.5  4.5 6.6  /

\setdashpattern <1.5mm, 0.5mm, 0.2mm, 0.5mm>
\setdashes

\plot  3.5  0.5  3.5 0.0  /
\plot  4.5  1.5  4.5 0.0  /

\plot  3.5  0.5  0.0 0.5  /
\plot  4.5  1.5  3.5 1.5  /

\plot  3.5  5.0  0.0 5.0  /
\plot  3.5  6.0  0.0 6.0  /
\plot  3.5  5.0  4.5 6.0  /

\put {$\bullet$} [c]  at 3.5 0.5
\put {$\bullet$} [c]  at 4.5 1.5
\put {$\bullet$} [c]  at 3.5 5.0
\put {$\bullet$} [c]  at 4.5 6.0

\put {$t^n$}     [c] at 3.5 -0.3
\put {$t^{n+1}$} [c] at 4.6 -0.3
\put {$h $}      [c] at 4.0  0.25
\put {$r^n$}     [c] at 3.8  1.1
\put {$f^n_k$}   [c] at 3.2  5.5

\put {${\rm df}^n(0)$}     [c] at 2.5  0.8
\put {${\rm df}^n(h)$}     [c] at 2.0  1.5
\put {${\rm df}^n(s_k)$}   [c] at 2.0  4.5
\put {${\rm df}^{n+1}(0)$} [l] at 4.7  1.5
\put {${\rm df}^{n+1}(s_k)$}     [l] at 4.7  6.0
\put {${\rm df}^{n}(s_k+h)$}     [c] at 3.5  6.9

\put {${\rm df}^0(t^{n+1}+s_k)$}   [l] at 0.4  6.4

\put {$y^n_k$}     [c] at 3.2  3.0
\put {$y^{n+1}_k$}     [l] at 4.6  4.0

\put {$T$}     [r] at -0.15  7.0
\put {$t$}     [l] at 10.2   0.0
\put {$s=T-t=0$}     [l] at  6.15   3

\setsolid
\arrow <1,5mm>   [0.25,0.75] from  3.1 0.8 to  3.3 0.65
\arrow <1,5mm>   [0.25,0.75] from  2.7 1.5 to  3.3 1.5
\arrow <1,5mm>   [0.25,0.75] from  2.7 4.5 to  3.3 4.9
\arrow <1,5mm>   [0.25,0.75] from  0.3 6.3 to  0.1 6.1
\arrow <1,5mm>   [0.25,0.75] from  3.1 6.7 to 3.4 6.1

\endpicture     \]
\caption{ The discretization of model equation for constructing probability field.}

\end{figure}


%% file: fig_401.tex
  

 


\begin{figure}[hbt] 

\[  \beginpicture   \setlinear 

\setcoordinatesystem units <10mm,10mm>
\setplotarea x from  -5.5 to 5.5, y from 0 to 6

\thinlines

\arrow <1,5mm>   [0.25,0.75] from  -5.0 0 to 5.0  0
\arrow <1,5mm>   [0.25,0.75] from  0.0 0.0 to  0.0 6.0

\plot  -2  1    2  1  /
\plot  -3  2.1  3  2.1  /
\plot  -4  3.5  4  3.5  /
\plot  -5  5  5  5  /

\plot 0 0   -2 1  -3  2.1     -4  3.5    -5  5 /
\plot 0 0    2 1   3  2.1      4  3.5     5  5 /

\plot 0 0  0.5 1   0.75  2.1   1 3.5  1.25 5 /
\plot 0 0  1.0 1   1.5   2.1   2 3.5  2.5 5 /
\plot 0 0  1.5 1   2.25  2.1   3 3.5  3.75 5 /

\plot 0 0  -0.5 1   -0.75  2.1   -1 3.5  -1.25 5 /
\plot 0 0  -1.0 1   -1.5   2.1   -2 3.5  -2.5 5 /
\plot 0 0  -1.5 1   -2.25  2.1   -3 3.5  -3.75 5 /

\put {0} [c] at 0 -0.22
\put {$x$} [c] at 5.0 0.3
\put {$\Delta x^1$} [c] at 2 0.5
\arrow <1,5mm>   [0.25,0.75] from 1.95 0.6 to 1.7 0.95

\put {$\Delta x^2$} [c] at 2.75 2.35
\put {$\Delta x^n$} [c] at 4.6 5.25

\put {$t$} [c] at 0.3 6.0

\plot -5 5.05  -5 5.6 /  
\arrow <1,5mm>   [0.25,0.75] from -2 5.4 to -5 5.4 
\arrow <1,5mm>   [0.25,0.75] from -2 5.4 to 0 5.4 

\put { 4$\cdot$StDev[$x(t^n)$]} [c] at -2.5 5.7

\endpicture     \]
\caption{A tree-like grid field.}

\end{figure}


%% file: fig_402.tex
  

 


\begin{figure}[hbt]

\[ \beginpicture    \setlinear

\setcoordinatesystem units <15mm,20mm>
\setplotarea x from -0.5 to 7.5, y from -0.2 to 1

\plot 0.0  0  5.7 0 / 
\arrow <1,5mm>   [0.25,0.75] from  5.8  0  to 7.1  0
\put {$x$} at 7.3 -0.01

\plot 0.0  0  0.0  -0.1 /
\plot 0.8  0  0.8  -0.1 /
\plot 1.6  0  1.6  -0.1 /
\plot 2.4  0  2.4  -0.1 /
\plot 3.2  0  3.2  -0.1 /
\plot 4.0  0  4.0  -0.1 /
\plot 4.8  0  4.8  -0.1 /
\plot 5.6  0  5.6  -0.1 /

\plot 0 1  7 1 /
\plot 0  1  0  0.9 /
\plot 1  1  1  0.9 /
\plot 2  1  2  0.9 /
\plot 3  1  3  0.9 /
\plot 4  1  4  0.9 /
\plot 5  1  5  0.9 /
\plot 6  1  6  0.9 /
\plot 7  1  7  0.9 /

\arrow <1,5mm>   [0.25,0.75] from  2.4  0  to 3.4 1
\arrow <1,5mm>   [0.25,0.75] from  2.4  0  to 4.7 1
\arrow <1,5mm>   [0.25,0.75] from  2.4  0  to 1.9 1
\arrow <1,5mm>   [0.25,0.75] from  2.4  0  to 0.7 1

\put {$x^{n-1}_i$}    at 2.4 -0.25
\put {$p^{n-1}_{i*}$} at 3.8  0.4
\put {$x^{n}_{*}$}    at 4.6  0.8

\put {$x^{n}_{j}$}    at 5.05  0.75
\put {$x^{n}_{j+1}$}  at 6.1   0.75
\put {$x^{n}_{j-1}$}  at 3.90   1.25
\put {$x^{n}_{j-2}$}  at 3.0    1.2

\plot 
4.9927 	1.0464 
4.9837 	1.0681 
4.9714 	1.0882 
4.9561 	1.1061 
4.9382 	1.1214 
4.9181 	1.1337 
4.8964 	1.1427 
4.8735 	1.1482 
4.8500 	1.1500 
4.8265 	1.1482 
4.8036 	1.1427 
4.7819 	1.1337 
4.7618 	1.1214 
4.7439 	1.1061 
4.7286 	1.0882 
4.7163 	1.0681 
4.7073 	1.0464 
4.7018 	1.0235   /
\arrow <1,5mm>   [0.45,0.75] from 4.9927 	1.0464  to 4.999 1.03

\plot
5.9920 	1.0508 
5.9682 	1.1004 
5.9292 	1.1475 
5.8759 	1.1910 
5.8096 	1.2298 
5.7321 	1.2629 
5.6451 	1.2896 
5.5509 	1.3091 
5.4517 	1.3210 
5.3500 	1.3250 
5.2483 	1.3210 
5.1491 	1.3091 
5.0549 	1.2896 
4.9679 	1.2629 
4.8904 	1.2298 
4.8241 	1.1910 
4.7708 	1.1475 
4.7318 	1.1004 
4.7080 	1.0508   /
\arrow <1,5mm>   [0.45,0.75] from 5.9960 1.04  to 5.999 1.03
\put {$p^{n}_{*,j+1}$} at 5.45 1.47

\plot
4.0043 	1.0329 
4.0171 	1.0649 
4.0381 	1.0953 
4.0668 	1.1234 
4.1025 	1.1485 
4.1443 	1.1699 
4.1911 	1.1871 
4.2418 	1.1997 
4.2952 	1.2074 
4.3500 	1.2100 
4.4048 	1.2074 
4.4582 	1.1997 
4.5089 	1.1871 
4.5557 	1.1699 
4.5975 	1.1485 
4.6332 	1.1234 
4.6619 	1.0953 
4.6829 	1.0649 
4.6957 	1.0329 /
\arrow <1,5mm>   [0.45,0.75] from 4.005 1.04  to 4.0 1.03

\arrow <1,5mm>   [0.25,0.75] from  0 1.1  to 0 1.9
\put {$t$} at 0 2.0

\put {$t^n$} at -0.3 1.0
\put {$t^{n-1}$} at -0.3 0.0

\endpicture     \]
\caption{ Transition probability between grid points.}

\end{figure}


%% file: fig_501.tex
  

 


\begin{figure}[hbt]

\[ \beginpicture    \setlinear

\setcoordinatesystem units <13mm,13mm>
\setplotarea x from -0.3 to 8.0, y from -0.5 to 5

\plot 0    0    0.7    0  /
\plot 0.6  -0.1    0.8    0.1  / 
\plot 0.7  -0.1    0.9    0.1  /
 
\arrow <1,5mm>   [0.25,0.75] from  0.8  0  to 7.1  0
\arrow <1,5mm>   [0.25,0.75] from  0.0  0  to 0    5

\plot  1.8  0  6.8  5.0 /

\plot  2  0.2   2   4.8 /
\plot  3  1.2   3   2.8 /
\plot  4  2.2   4   3.8 /
\plot  5  3.2   5   4.8 /

\put {$\bullet$} [c]  at  2  0.2
\put {$\bullet$} [c]  at  3  1.2
\put {$\bullet$} [c]  at  4  2.2
\put {$\bullet$} [c]  at  5  3.2

\put {$\bullet$} [c]  at  2  1.2
\put {$\bullet$} [c]  at  2  2.2
\put {$\bullet$} [c]  at  2  3.2
\put {$\bullet$} [c]  at  2  4.2

\put {$\bullet$} [c]  at  3  2.2
\put {$\bullet$} [c]  at  4  3.2
\put {$\bullet$} [c]  at  5  4.2

\put {$T$}   [c]  at  0  5.2
\put {$t^n$} [c]  at  2 -0.3
\put {$t^\alpha$} [c]  at  3 -0.3
\put {$t^\beta$}  [c]  at  4 -0.3
\put {$t^\gamma$} [c]  at  5 -0.3

\put {$c^n_1$} [c]  at  1.8 0.7
\put {$c^n_2$} [c]  at  1.8 1.7
\put {$c^n_3$} [c]  at  1.8 2.7
\put {$c^n_4$} [c]  at  1.8 3.7

\put {$c^\alpha_1$} [c]  at  2.8 1.6
\put {$c^\beta_1$} [c]   at  3.8 2.6
\put {$c^\gamma_1$} [c]  at  4.8 3.6

\put {${\rm df}^n(s)$} [c]  at  2 5
\put {${\rm df}^\alpha(s)$} [c]  at  3  3
\put {${\rm df}^\beta(s)$}  [c]  at  4  4
\put {${\rm df}^\gamma(s)$} [c]  at  5  5

\put {$s=T-t=0$} [l]  at  5.5 3.5
\put {$t$} [l]  at  7.2 -0.1

\setdashes

\plot  3  1.2   3   0 /
\plot  4  2.2   4   0 /
\plot  5  3.2   5   0 /

\endpicture     \]
\caption{ A sketch for cash flow calculation.}

\end{figure}


%% file: fig_601.tex
  

 


\begin{figure}[hbt]

\[ \beginpicture    \setlinear

\setcoordinatesystem units <10mm,8mm>
\setplotarea x from 0.0 to 8.0, y from 0 to 6

\arrow <1,5mm>   [0.25,0.75] from  0.0  6  to 8  6
\arrow <1,5mm>   [0.25,0.75] from  0.0  6  to 0  0

\plot  2  1  2  5.0 /
\plot  4  1  4  5.0 /
\plot  6  1  6  5.0 /

\plot  2  1  6  1 /
\plot  2  3  6  3 /
\plot  2  5  6  5 /

\setdashes

\plot  2  1  2  6 /
\plot  4  1  4  6 /
\plot  6  1  6  6 /

\plot  2  1  0  1 /
\plot  2  3  0  3 /
\plot  2  5  0  5 /

\put {1}     [c] at 3.8 2.7
\put {2}     [c] at 4.0 0.7
\put {3}     [c] at 6.2 0.8

\put {4}     [c] at 6.2 3
\put {5}     [c] at 6.2 5.2
\put {6}     [c] at 4.25 5.35

\put {7}     [c] at 1.8 5.35
\put {8}     [c] at 1.8 3.35
\put {9}     [c] at 1.9 0.7

\put {$A$}   [c] at 5 2
\put {$B$}   [c] at 5 4
\put {$C$}   [c] at 3 4
\put {$D$}   [c] at 3 2

\put {$t$}   [c] at -0.4 0
\put {$t^n$}   [c] at -0.4 3
\put {$t^{n+1}$}   [c] at -0.4 1
\put {$\tau$}   [c] at 0.2 2

\put {$s$}         [c] at 8.1 6.3
\put {$s_j$}       [c] at 4 6.3
\put {$s_{j+1}$}   [c] at 6 6.3
\put {$\delta$}    [c] at 5 5.7

\endpicture     \]
\caption{ Finite element mesh for volatility calibration. }

\end{figure}


%% file: table_801.tex
  

 


\begin{table}[hbt]
\begin{center}
\caption{USD LIBOR/OIS swap rates and lognormal volatility on 2015-08-03.}
\vspace{4mm}
\begin{small}
\begin{tabular}{|c|c|c|c|c|c|c|c|c|c|c|} \hline
\tt	LIBOR	&\tt	0.545	&\tt	0.904	&\tt	1.219	&\tt	1.473	&\tt	1.683	&\tt	1.997	&\tt	2.281	&\tt	2.522	&\tt	2.631	&\tt	2.710	\\
\tt	OIS	&\tt	0.385	&\tt	0.708	&\tt	1.005	&\tt	1.250	&\tt	1.450	&\tt	1.763	&\tt	2.043	&\tt	2.288	&\tt	2.400	&\tt	2.485	\\
\hline
\tt	Vol	&\tt	1Y	&\tt	2Y	&\tt	3Y	&\tt	4Y	&\tt	5Y	&\tt	7Y	&\tt	10Y	&\tt	15Y	&\tt	20Y	&\tt	30Y	\\
\hline
\tt	3M	&\tt	69.63	&\tt	60.75	&\tt	52.34	&\tt	48.34	&\tt	44.53	&\tt	38.72	&\tt	34.49	&\tt	31.23	&\tt	29.69	&\tt	28.52	\\
\tt	6M	&\tt	64.58	&\tt	57.36	&\tt	52.02	&\tt	47.53	&\tt	43.77	&\tt	38.40	&\tt	34.65	&\tt	31.62	&\tt	30.06	&\tt	28.90	\\
\tt	1Y	&\tt	58.03	&\tt	52.62	&\tt	47.96	&\tt	44.38	&\tt	41.38	&\tt	37.07	&\tt	33.87	&\tt	31.17	&\tt	29.68	&\tt	28.56	\\
\tt	2Y	&\tt	49.78	&\tt	45.92	&\tt	42.16	&\tt	39.31	&\tt	37.25	&\tt	34.34	&\tt	32.06	&\tt	29.73	&\tt	28.55	&\tt	27.53	\\
\tt	3Y	&\tt	43.88	&\tt	40.15	&\tt	37.71	&\tt	35.93	&\tt	34.58	&\tt	32.75	&\tt	30.76	&\tt	28.57	&\tt	27.58	&\tt	26.74	\\
\tt	4Y	&\tt	39.24	&\tt	36.21	&\tt	34.73	&\tt	33.60	&\tt	32.67	&\tt	31.24	&\tt	29.81	&\tt	27.67	&\tt	26.61	&\tt	25.86	\\
\tt	5Y	&\tt	35.75	&\tt	34.08	&\tt	33.11	&\tt	32.27	&\tt	31.51	&\tt	30.32	&\tt	29.12	&\tt	27.03	&\tt	25.99	&\tt	25.27	\\
\tt	7Y	&\tt	31.68	&\tt	30.88	&\tt	30.54	&\tt	29.86	&\tt	29.28	&\tt	28.42	&\tt	27.38	&\tt	25.38	&\tt	24.40	&\tt	23.78	\\
\tt	10Y	&\tt	27.88	&\tt	27.29	&\tt	27.02	&\tt	26.96	&\tt	26.67	&\tt	26.07	&\tt	25.20	&\tt	23.33	&\tt	22.35	&\tt	21.87	\\
\hline
\end{tabular}
\end{small}
\end{center}
\end{table}


%% file: fig_801.tex
  

 


\begin{figure}[hbt]

\[ \beginpicture    \setlinear

\setcoordinatesystem units <5mm,0.95mm>
\setplotarea x from 0 to 20, y from 0 to 70
\axis bottom label {Exercise Period \#} shiftedto y=0  
		ticks numbered from 0 to 20 by 5 unlabeled short from 1 to 19 by 1 /
\axis left label  {USD}
		ticks numbered from 0 to 70  by 10 unlabeled short from 5 to 65 by 5 /

\plot
1	0.00 
2	-0.56 
3	1.20 
4	8.90 
5	15.59 
6	20.48 
7	29.25 
8	35.58 
9	39.41 
10	48.34 
11	50.15 
12	51.42 
13	53.48 
14	58.57 
15	56.58 
16	59.70 
17	59.61 
18	61.38 
19	60.48  /
\plot  10.5 37 12 37 /
\put { This Model MC/1F } [l]  at 12.2 37

\setdashes
\plot
1	0.19 
2	-0.24 
3	1.73 
4	10.05 
5	15.59 
6	19.71 
7	29.81 
8	33.73 
9	38.95 
10	49.12 
11	49.81 
12	50.16 
13	53.48 
14	59.15 
15	56.10 
16	60.70 
17	59.56 
18	61.64 
19	60.51  /
\plot  10.5 29 12 29 /
\put { This Model MC/3F } [l]  at 12.2 29

\setdots
\setdashpattern <1.5mm, 0.7mm, 0.2mm, 0.7mm>

\plot
1	0.00 
2	0.87 
3	4.72 
4	10.46 
5	15.50 
6	23.34 
7	29.39 
8	35.60 
9	42.97 
10	48.28 
11	47.75 
12	52.58 
13	52.19 
14	55.61 
15	55.60 
16	56.20 
17	55.70 
18	56.54 
19	54.47  /
\plot  10.5 21 12 21 /
\put { This Model Grid/1F } [l]  at 12.2 21

\multiput {$\circ$} at
1	2.00 
2	0.85 
3	4.29 
4	10.26 
5	16.48 
6	23.54 
7	28.59 
8	34.96 
9	39.40 
10	44.94 
11	47.18 
12	52.32 
13	51.85 
14	56.00 
15	55.28 
16	56.68 
17	56.68 
18	58.39 
19	57.08  /
\put { $\circ$ } [c] at 11.25 13
\put { BDT/HW Grid/1F } [l]  at 12.2 13

\multiput {$\bullet$} at
1	0.00 
2	0.43 
3	3.16 
4	8.71 
5	14.69 
6	21.93 
7	27.73 
8	34.89 
9	39.46 
10	44.93 
11	47.11 
12	52.49 
13	52.22 
14	55.12 
15	54.23 
16	55.71 
17	56.15 
18	58.24 
19	57.02  /
\put { $\bullet$ } [c] at 11.25 5
\put { 2-Plus Grid/3F } [l]  at 12.2 5

\put {10-Year Bermudan Swaption } [l] at 2 60
\put {Option Value} [l] at  0.4 70

\endpicture     \]
\caption{ The value distribution for a Bermudan option to enter a 10-year break-even swap. }

\end{figure}


%% file: fig_802.tex
  

 


\begin{figure} 

\[ \beginpicture    \setlinear

\setcoordinatesystem units <5mm,2.3mm>
\setplotarea x from 0 to 20, y from 0 to 30
\axis bottom label {Exercise Period \#} shiftedto y=0  
		ticks numbered from 0 to 20 by 5 unlabeled short from 1 to 19 by 1 /
\axis left label  {USD}
		ticks numbered from 0.0 to 30  by 5 unlabeled short from 2.5 to 27.5 by 5.0 /

\plot
1	1.11 
2	6.73 
3	7.40 
4	14.00 
5	14.33 
6	13.72 
7	18.14 
8	17.68 
9	19.32 
10	25.66 
11	23.29 
12	22.94 
13	25.25 
14	26.06 
15	26.53 
16	27.21 
17	27.81 
18	28.47 
19	29.34 
  /
\plot  10.5 16.5 12 16.5 /
\put { This Model MC/1F } [l]  at 12.2 16.5

\setdashes
\plot
1	1.19 
2	5.26 
3	7.74 
4	12.96 
5	13.35 
6	13.87 
7	19.49 
8	17.21 
9	20.74 
10	25.39 
11	22.93 
12	22.10 
13	23.08 
14	26.58 
15	25.71 
16	28.10 
17	27.89 
18	28.53 
19	29.36 
 /
\plot  10.5 13 12 13 /
\put { This Model MC/3F } [l]  at 12.2 13

\setdots
\setdashpattern <1.5mm, 0.7mm, 0.2mm, 0.7mm>

\plot
1	1.04 
2	6.85 
3	12.88 
4	14.73 
5	15.46 
6	16.53 
7	17.77 
8	20.25 
9	20.17 
10	26.01 
11	21.66 
12	22.71 
13	23.34 
14	24.18 
15	24.65 
16	24.02 
17	24.34 
18	23.43 
19	23.33  /
\plot  10.5 9.5 12 9.5 /
\put { This Model Grid/1F } [l]  at 12.2 9.5

\multiput {$\circ$} at
1	2.71 
2	7.15 
3	10.04 
4	11.77 
5	14.02 
6	15.44 
7	16.80 
8	17.60 
9	18.94 
10	20.01 
11	21.30 
12	22.23 
13	23.00 
14	23.96 
15	24.77 
16	24.43 
17	25.03 
18	25.33 
19	25.93  /
\put { $\circ$ } [c] at 11.25 6
\put { BDT/HW Grid/1F } [l]  at 12.2 6

\multiput {$\bullet$} at
1	2.42 
2	7.22 
3	10.12 
4	12.15 
5	14.45 
6	16.01 
7	17.78 
8	18.93 
9	20.10 
10	20.90 
11	22.14 
12	23.61 
13	23.61 
14	24.51 
15	24.43 
16	23.55 
17	24.66 
18	25.14 
19	25.87  /
\put { $\bullet$ } [c] at 11.25 2.5
\put { 2-Plus Grid/3F } [l]  at 12.2 2.5

\put {10-Year Callble Swap } [l] at 5 27
\put {Option Value} [l] at  0.4 30

\endpicture     \]
\caption{ The value distribution for a Bermudan option to cancel a 10-year break-even swap. }

\end{figure}


%% file: fig_803.tex
  

 


\begin{figure}[hbt]

\[ \beginpicture    \setlinear

\setcoordinatesystem units <5mm,0.22mm>
\setplotarea x from 0 to 20, y from -300 to 50
\axis bottom label {Exercise Period \#} shiftedto y=-300  
		ticks numbered from 0 to 20 by 5 unlabeled short from 1 to 19 by 1 /
\axis left label  {\$} ticks numbered from -300 to 50  by 50  /

\plot
1	-258.57 
2	-225.36 
3	-189.50 
4	-153.81 
5	-121.28 
6	-98.48 
7	-77.52 
8	-63.79 
9	-46.66 
10	-34.89 
11	-23.38 
12	-17.78 
13	-9.89 
14	-8.26 
15	-3.79 
16	-3.96 
17	-0.40 
18	-1.20 
19	0.56 
20	-0.41 
/

\multiput {$\bullet$} at 
1	16.77675
2	-12.1645
3	-43.74275
4	-69.19075
5	-77.10675
6	-106.48075
7	-105.40475
8	-110.479
9	-127.7235
10	-120.236
11	-135.31075
12	-120.73475
13	-120.317
14	-124.1465
15	-112.6365
16	-112.0875
17	-98.83525
18	-87.31175
19	-86.21825 
/
\plot  7 -190 8.8 -190 /
\put { $\bullet$ } [c] at 6.5 -190
\put { This Model MC/1F } [l]  at 8.7 -190


\setdashes
\plot
1	-258.10 
2	-215.73 
3	-162.92 
4	-116.70 
5	-80.46 
6	-56.70 
7	-36.51 
8	-25.04 
9	-10.72 
10	-2.47 
11	7.94 
12	11.40 
13	19.12 
14	19.13 
15	22.73 
16	21.37 
17	24.09 
18	23.61 
19	25.27
20   23.79 
/

\multiput {$\circ$} at 
1	-25.035
2	-79.45175
3	-105.591
4	-121.473
5	-123.59925
6	-136.5345
7	-107.74425
8	-119.61775
9	-123.70225
10	-132.44875
11	-148.1735
12	-137.64975
13	-145.049
14	-135.439
15	-132.24375
16	-132.30175
17	-123.318
18	-112.61925
19	-110.53375
/

\plot  7 -220 8.8 -220 /
\put { $\circ$ } [c] at 6.5 -220
\put { This Model MC/3F } [l]  at 8.7 -220


\setdashpattern <1.9mm, 0.7mm, 0.2mm, 0.7mm>
\plot
1	-258.04 
2	-225.23 
3	-189.18 
4	-153.26 
5	-120.72 
6	-98.60 
7	-77.87 
8	-64.43 
9	-48.90 
10	-38.73 
11	-27.71 
12	-22.00 
13	-13.86 
14	-11.26 
15	-7.05 
16	-8.41 
17	-5.83 
18	-7.04 
19	-4.86 
20	-5.81 
/

\multiput {$*$} at 
1	-152.5738916
2	-133.3336314
3	-127.1166967
4	-124.1011215
5	-118.6739362
6	-116.7194142
7	-115.3211606
8	-114.8507648
9	-113.0527363
10	-111.6326676
11	-110.2840655
12	-106.4537132
13	-109.7112292
14	-100.4716728
15	-97.03623757
16	-94.38762605
17	-92.03722588
18	-91.58739378
19	-91.81845537
/

\plot  7 -250 8.8 -250 /
\put { $*$ } [c] at 6.5 -250
\put { BDT/HW Grid/1F } [l]  at 8.7 -250


\setdots
\plot
1	-258.04 
2	-225.23 
3	-189.18 
4	-153.25 
5	-120.39 
6	-97.60 
7	-75.46 
8	-59.36 
9	-41.23 
10	-29.50 
11	-17.25 
12	-12.12 
13	-4.78 
14	-3.44 
15	0.52 
16	-0.01 
17	3.54 
18	2.84 
19	4.63 
20	2.88 
/

\multiput { \tiny $\times$} at 
1	-157.71662
2	-144.5931317
3	-143.7781707
4	-144.1368994
5	-139.7586615
6	-136.4517873
7	-130.5686932
8	-127.6093454
9	-124.6491189
10	-122.2698535
11	-118.0053453
12	-109.2165679
13	-112.0256495
14	-108.5522281
15	-113.397573
16	-113.5364673
17	-109.260484
18	-104.4083595
19	-100.0180357
/

\plot  7 -280 8.8 -280 /
\put {\tiny $\times$ } [c] at 6.5 -280
\put { 2-Plus Grid/3F } [l]  at 8.7 -280

\setsolid
\thinlines

\plot 20 -300 20 50 /
\plot 20 -300 20.4 -300 /
\plot 20 -250 20.4 -250 /
\plot 20 -200 20.4 -200 /
\plot 20 -150 20.4 -150 /
\plot 20 -100 20.4 -100 /
\plot 20 -50 20.4 -50 /
\plot 20   0 20.4   0 /
\plot 0  50 20.4  50 /

\put { 0 }  [l]   at 20.5 -300
\put { 20 } [l]   at 20.5 -250
\put { 40 } [l]   at 20.5 -200
\put { 60 } [l]   at 20.5 -150
\put { 80 } [l]   at 20.5 -100
\put { 100 } [l]  at 20.5 -50
\put { 120 } [l]  at 20.5 0
\put { 140 } [l]  at 20.5 50

\put { Swap Value } [c]  at  9 20
\put { Option Value } [l]  at  12 -80

\put { Swap } [c]  at  0  70
\put { Option } [c]  at  20 70

\endpicture     \]
\caption{ The distribution of swap and option values in a callable CMS spread range-accrual swap deal. }

\end{figure}


%% file: table_802.tex
 

 


\begin{table}[hbt]
\begin{center}
\caption{Bucket Vegas of a Bermudan option to enter a 10-year swap.}
\vspace{4mm}
\begin{small}
\begin{tabular}{|c|r|r|r|r|r|r|r|r|r|r|} \hline
\tt	Vega	&\tt	1Y	&\tt	2Y	&\tt	3Y	&\tt	4Y	&\tt	5Y	&\tt	6Y	&\tt	7Y	&\tt	8Y	&\tt	9Y	&\tt	10Y	\\
\hline
\tt	1Y	&\tt	-7 	&\tt	-5 	&\tt	-1 	&\tt	3 	&\tt	9 	&\tt	17 	&\tt	-8 	&\tt	-14 	&\tt	467 	&\tt	0 	\\
\tt	2Y	&\tt	26 	&\tt	20 	&\tt	9 	&\tt	0 	&\tt	-14 	&\tt	-40 	&\tt	14 	&\tt	399 	&\tt	0 	&\tt	0 	\\
\tt	3Y	&\tt	-5 	&\tt	2 	&\tt	11 	&\tt	-39 	&\tt	-39 	&\tt	-34 	&\tt	261 	&\tt	0 	&\tt	0 	&\tt	0 	\\
\tt	4Y	&\tt	-26 	&\tt	-63 	&\tt	-79 	&\tt	55 	&\tt	73 	&\tt	202 	&\tt	0 	&\tt	0 	&\tt	0 	&\tt	0 	\\
\tt	5Y	&\tt	-22 	&\tt	56 	&\tt	70 	&\tt	-35 	&\tt	66 	&\tt	0 	&\tt	0 	&\tt	0 	&\tt	0 	&\tt	0 	\\
\tt	6Y	&\tt	40 	&\tt	-13 	&\tt	-24 	&\tt	61 	&\tt	0 	&\tt	0 	&\tt	0 	&\tt	0 	&\tt	0 	&\tt	0 	\\
\tt	7Y	&\tt	-13 	&\tt	-57 	&\tt	25 	&\tt	0 	&\tt	0 	&\tt	0 	&\tt	0 	&\tt	0 	&\tt	0 	&\tt	0 	\\
\tt	8Y	&\tt	-13 	&\tt	60 	&\tt	0 	&\tt	0 	&\tt	0 	&\tt	0 	&\tt	0 	&\tt	0 	&\tt	0 	&\tt	0 	\\
\tt	9Y	&\tt	13 	&\tt	0 	&\tt	0 	&\tt	0 	&\tt	0 	&\tt	0 	&\tt	0 	&\tt	0 	&\tt	0 	&\tt	0 	\\
\tt	10Y	&\tt	0 	&\tt	0 	&\tt	0 	&\tt	0 	&\tt	0 	&\tt	0 	&\tt	0 	&\tt	0 	&\tt	0 	&\tt	0 	\\
\hline
\end{tabular}
\end{small}
\end{center}
\end{table}


%% file: table_803.tex
  

 


\begin{table}[hbt]
\begin{center}
\caption{Bucket Vegas of a Bermudan option to cancel a 10-year swap.}
\vspace{4mm}
\begin{small}
\begin{tabular}{|c|r|r|r|r|r|r|r|r|r|r|} \hline
\tt	Vega	&\tt	1Y	&\tt	2Y	&\tt	3Y	&\tt	4Y	&\tt	5Y	&\tt	6Y	&\tt	7Y	&\tt	8Y	&\tt	9Y	&\tt	10Y	\\
\hline
\tt	1Y	&\tt	-7 	&\tt	-5 	&\tt	-1 	&\tt	4 	&\tt	10 	&\tt	20 	&\tt	-8 	&\tt	-15 	&\tt	256 	&\tt	0 	\\
\tt	2Y	&\tt	26 	&\tt	20 	&\tt	10 	&\tt	-2 	&\tt	-15 	&\tt	-43 	&\tt	16 	&\tt	554 	&\tt	0 	&\tt	0 	\\
\tt	3Y	&\tt	-8 	&\tt	-7 	&\tt	-1 	&\tt	-31 	&\tt	-31 	&\tt	-30 	&\tt	359 	&\tt	0 	&\tt	0 	&\tt	0 	\\
\tt	4Y	&\tt	-34 	&\tt	-33 	&\tt	-47 	&\tt	36 	&\tt	48 	&\tt	236 	&\tt	0 	&\tt	0 	&\tt	0 	&\tt	0 	\\
\tt	5Y	&\tt	-5 	&\tt	16 	&\tt	28 	&\tt	-40 	&\tt	61 	&\tt	0 	&\tt	0 	&\tt	0 	&\tt	0 	&\tt	0 	\\
\tt	6Y	&\tt	5 	&\tt	29 	&\tt	3 	&\tt	49 	&\tt	0 	&\tt	0 	&\tt	0 	&\tt	0 	&\tt	0 	&\tt	0 	\\
\tt	7Y	&\tt	54 	&\tt	-207 	&\tt	-5 	&\tt	0 	&\tt	0 	&\tt	0 	&\tt	0 	&\tt	0 	&\tt	0 	&\tt	0 	\\
\tt	8Y	&\tt	-73 	&\tt	178 	&\tt	0 	&\tt	0 	&\tt	0 	&\tt	0 	&\tt	0 	&\tt	0 	&\tt	0 	&\tt	0 	\\
\tt	9Y	&\tt	32 	&\tt	0 	&\tt	0 	&\tt	0 	&\tt	0 	&\tt	0 	&\tt	0 	&\tt	0 	&\tt	0 	&\tt	0 	\\
\tt	10Y	&\tt	0 	&\tt	0 	&\tt	0 	&\tt	0 	&\tt	0 	&\tt	0 	&\tt	0 	&\tt	0 	&\tt	0 	&\tt	0 	\\
\hline
\end{tabular}
\end{small}
\end{center}
\end{table}


%% file: table_804.tex
  

 


\begin{table}[hbt]
\begin{center}
\caption{ Distribution of bucket Vegas on a CMS spread swap.}
\vspace{4mm}
\begin{small}
\begin{tabular}{|c|r|r|r|r|r|r|r|r|r|r|} \hline
\tt	Vega	&\tt	1Y	&\tt	2Y	&\tt	3Y	&\tt	4Y	&\tt	5Y	&\tt	6Y	&\tt	7Y	&\tt	8Y	&\tt	9Y	&\tt	10Y	\\
\hline
\tt	1Y	&\tt	0 	&\tt	0 	&\tt	0 	&\tt	0 	&\tt	0 	&\tt	0 	&\tt	0 	&\tt	0 	&\tt	0 	&\tt	1 	\\
\tt	2Y	&\tt	0 	&\tt	0 	&\tt	0 	&\tt	0 	&\tt	0 	&\tt	0 	&\tt	0 	&\tt	0 	&\tt	-1 	&\tt	-3 	\\
\tt	3Y	&\tt	-2 	&\tt	-11 	&\tt	2 	&\tt	2 	&\tt	3 	&\tt	4 	&\tt	4 	&\tt	4 	&\tt	7 	&\tt	51 	\\
\tt	4Y	&\tt	-4 	&\tt	-23 	&\tt	1 	&\tt	3 	&\tt	7 	&\tt	10 	&\tt	12 	&\tt	15 	&\tt	12 	&\tt	107 	\\
\tt	5Y	&\tt	-2 	&\tt	-3 	&\tt	2 	&\tt	4 	&\tt	4 	&\tt	2 	&\tt	3 	&\tt	3 	&\tt	8 	&\tt	38 	\\
\tt	6Y	&\tt	0 	&\tt	0 	&\tt	0 	&\tt	0 	&\tt	0 	&\tt	0 	&\tt	0 	&\tt	0 	&\tt	0 	&\tt	0 	\\

\hline
\end{tabular}
\end{small}
\end{center}
\end{table}


%% file: table_805.tex
  

 


\begin{table}[hbt]
\begin{center}
\caption{ Distribution of bucket Vegas on a callable CMS spread swap.}
\vspace{4mm}
\begin{small}
\begin{tabular}{|c|r|r|r|r|r|r|r|r|r|r|} \hline
\tt	Vega	&\tt	1Y	&\tt	2Y	&\tt	3Y	&\tt	4Y	&\tt	5Y	&\tt	6Y	&\tt	7Y	&\tt	8Y	&\tt	9Y	&\tt	10Y	\\
\hline
\tt	1Y	&\tt	0 	&\tt	1 	&\tt	0 	&\tt	0 	&\tt	0 	&\tt	0 	&\tt	0 	&\tt	0 	&\tt	0 	&\tt	-2 	\\
\tt	2Y	&\tt	0 	&\tt	-6 	&\tt	0 	&\tt	0 	&\tt	0 	&\tt	0 	&\tt	0 	&\tt	0 	&\tt	0 	&\tt	8 	\\
\tt	3Y	&\tt	0 	&\tt	150 	&\tt	-1 	&\tt	1 	&\tt	1 	&\tt	3 	&\tt	5 	&\tt	7 	&\tt	1 	&\tt	-182 	\\
\tt	4Y	&\tt	-5 	&\tt	292 	&\tt	8 	&\tt	7 	&\tt	5 	&\tt	-3 	&\tt	-3 	&\tt	-3 	&\tt	13 	&\tt	-321 	\\
\tt	5Y	&\tt	2 	&\tt	98 	&\tt	-3 	&\tt	-3 	&\tt	-3 	&\tt	6 	&\tt	7 	&\tt	8 	&\tt	-3 	&\tt	-104 	\\
\tt	6Y	&\tt	0 	&\tt	0 	&\tt	0 	&\tt	0 	&\tt	0 	&\tt	0 	&\tt	0 	&\tt	0 	&\tt	0 	&\tt	0 	\\
\hline
\end{tabular}
\end{small}
\end{center}
\end{table}
